\documentclass[]{interact}

\usepackage{graphicx}
\usepackage{amsmath,bm}
\usepackage[utf8x]{inputenc} 
\usepackage{float} 
\usepackage{datetime}
\usepackage{cite} 
\usepackage[numbers,sort&compress]{natbib}
\bibpunct[, ]{[}{]}{,}{n}{,}{,}
\makeatletter
\def\NAT@def@citea{\def\@citea{\NAT@separator}}
\makeatother

\usepackage{color,verbatim,epsfig}
\usepackage{amssymb,amsthm}

\newcommand{\blue}[1]{{\color{black}{#1}}}

\def\p{\partial}

\def\eps{\varepsilon}

\newcommand{\be}{\begin{equation}}

\newcommand{\ee}{\end{equation}}

\newcommand{\ov}{\overline{ v}}

\newcommand{\os}{\overline{{ s}}}

\newcommand{\bx}{{\bf x}}
\newcommand{\by}{{\bf y}}
\newcommand{\bv}{{\bf v}}
\newcommand{\bF}{{\bf f}}

\newcommand{\br}{{\bf r}}
\newcommand{\bX}{{\bf X}}

\newcommand{\RL}{{\rm Re}}

\begin{document}

\title{ 
Multiscale properties of Large Eddy Simulations: \\ 
correlations between resolved-scale 
velocity-field increments and subgrid-scale quantities \footnote{postprint version of the manuscript published in Journal of Turbulence (2018)}}

\author{
\name{Moritz Linkmann$^{1,2,*}$, Michele Buzzicotti$^1$ and Luca Biferale$^1$}
\affil{$^1$Dept. of Physics and INFN, University of Rome Tor Vergata, Rome, Italy \\
$^2$Fachbereich Physik, Philipps-Universit\"at Marburg, Marburg, Germany}
$^*$\email{moritz.linkmann@physik.uni-marburg.de}}

\maketitle
\begin{abstract}
We provide analytical and numerical results
concerning 
multi-scale correlations between the  resolved velocity field  and the subgrid-scale (SGS) stress-tensor in large eddy simulations (LES). Following previous studies for Navier-Stokes equations (NSE),  we derive the exact  hierarchy of LES equations governing the spatio-temporal evolution of velocity structure functions of any order. The aim is to assess the influence of the sub-grid model on the inertial range intermittency. 
We provide a series of predictions, within the multifractal theory,  for the scaling  of  correlation involving the SGS stress  and we  compare them against numerical results from high-resolution Smagorinsky LES 
  and from {\em a-priori} filtered data generated from direct numerical simulations (DNS).   We find that  LES data generally agree very well with 
filtered DNS results and with the multifractal prediction for all leading terms in the balance equations. Discrepancies are measured for some of the subleading terms involving cross-correlation between resolved velocity increments and the SGS tensor or the SGS energy transfer, suggesting that there must be room to improve the SGS modelisation to further extend the inertial range properties for any fixed LES resolution.
\end{abstract}


\section{Introduction}
One of the main challenges in numerical and experimental turbulence is the existence of
anomalously strong non-Gaussian fluctuations,
which are a generic feature of all three-dimensional flows \cite{Frisch95,Pope00,Lesieur08}. Such extreme
events occur in a variety of flow configurations, both on Eulerian and Lagrangian domains
\cite{Arneodo_et_al1996,Benzi99,antonia2000,Gotoh02,Qian02,desilva2015} and
become more and more important with increasing Reynolds number, $Re = U_0L_0/\nu$, where $U_0,L_0$ are the characteristic velocity and length scale of the flow, while $\nu$ is the viscosity.
The Reynolds number measures the relative importance of linear {\it vs}  non-linear terms in the Navier-Stokes evolution. For large $Re$, the dynamics becomes fully turbulent and an inertial-range energy cascade develops.  Power laws  with anomalous scaling exponents are observed  for moments of velocity increments  in the inertial range, a phenomenon known as {\it 
  intermittency}. No systematic derivation of the value of the scaling exponents is known from first principle, and the problem is considered key for both fundamental aspects  and its applied consequences, being connected to the existence of wild fluctuations in the velocity increments and in the  energy dissipation field. Empirical data are always affected by spurious and/or sub-leading contributions, making an accurate determination of the scaling exponents difficult. Hence, it is mandatory to develop more and more refined experimental and numerical techniques to increase the scaling range and/or to improve the scaling properties. State-of-the-art data in the laboratories reach a maximum inertial range extensions of one/two decades 
\cite{sinhuber2017}, and the exponents are often evaluated using
sophisticated finite-size-techniques as Extended Self Similarity \cite{benzi1993extended} in order to reduce spurious effects. 
Similarly, concerning numerical studies, despite the huge progresses  made in recent years 
\cite{benzi2010inertial,ishihara2009study,iyer2017reynolds} 
we are  still far from reaching a resolution high enough to give a firm statement about scaling, in particular concerning subtle issues connected to the alleged different statistical properties of longitudinal and transverse velocity increments, or of the enstrophy and stress.  

A potential alternative strategy to minimise viscous effects and to concentrate only on high Reynolds number properties is provided by  the application of  large eddy
simulation (LES), where we introduce a model for the  small-scale dynamics while fully resolving the most energy-containing scales 
\cite{Smagorinsky63,Deardorff70,Germano91,MeneveauARFM,Pope00,Sagaut06,Sagaut08,Lesieur08,lesieur2005large,leonard1975energy,khanna1997analysis,domaradzki1993analysis,domaradzki1997subgrid}. 
In this paper, we perform a first step in  order to assess how much LES can be used to
estimate inertial range scaling properties of fully developed turbulence. The aim being twofold, first to have a tool able to minimise viscous and small-scale effects on the inertial range, second to assess 
the performance of high Reynolds  LES {\it tout-court}  owing to the emergent
role of  high-resolution modelling  where the cutoff scale lies in the inertial
subrange  \cite{Stevens14,MartinezTossas16}.\\

In order to assess the performance of LES models in reproducing
the aforementioned extreme events with reasonable accuracy, it is first  mandatory   to 
understand the statistical coupling between the resolved velocity field 
and the subgrid model.  
The present paper is mainly concerned with this  point.  
To do that, we derive the  exact hierarchy of 
equations satisfied by  the generic $n$th order structure functions 
made in terms of moments of the resolved velocity increments and involving the correlations with the modelled subgrid-scale (SGS) stress-tensor. 
Furthermore, we provide a set of multifractal (MF) predictions  for the scaling behavior of 
all  correlations entering in the equations of motion, which are subsequently 
compared to data obtained from {\em a-priori} filtered direct numerical simulations (DNS) of homogeneous isotropic 
turbulence on up to $2048^3$ grid points and from {\it a-posteriori} highly resolved Smagorinsky LES using up 
to $1024^3$ grid points.  
\blue{We focus here on the Smagorinsky model party because of its simplicity and wide usage.
More importantly, the Smagorinsky model is unable to model interactions leading to
backscatter events. As most LES models include a dissipative part to prevent numerical
instabilities, the modelling of backscatter events is still a challenge in LES.
In view of potential applicability of LES models to study inertial-range physics,
we also wish to assess if and how the absence of backscatter
affects the scaling of the correlation functions, and the Smagorinsky model 
is particularly well suited to this part of the analysis.}
\\
The main conclusion is that already the Smagorinsky LES modelling is a good tool to minimise effects induced by the ultraviolet, large wavenumber, cut-off on the inertial range: all leading scaling properties measured on the real {\it a-priori} data are well reproduced by the {\it a-posteriori} LES data. This opens the way to perform highly resolved LES to improve the actual knowledge of the inertial range physics, by further minimising the dissipative effects. 
For the sake of simplicity, we start here to address only  homogeneous and isotropic turbulence but the whole machinery can be reproduced for bounded flows as well, at the price of a higher analytical complexity. 
\\
This paper is organised as follows. 
The structure function hierarchies are derived in 
Sec.~\ref{sec:les-hierarchy} and in Appendix~\ref{app:derivation} and \ref{app:ples-hierarchy} for different formulations of the filtered Navier-Stokes equations (NSE). Section \ref{sec:scaling} is concerned with the predictions for scaling behaviour of multi-points correlation functions based on 
the MF hypothesis. The numerical results are presented in Sec.~\ref{sec:numerics}, and 
we conclude with a summary  in Sec.~\ref{sec:conclusions}.   

\section{Structure function hierarchies for LES}
\label{sec:les-hierarchy}
The application of LES requires a splitting into resolved scales and unresolved (subgrid) scales.  
The resolved-scale quantities are defined though the application of a filter kernel 
$G^\Delta$ at a given scale scale $\Delta$
to the velocity field $\bv$
\be
\overline{\bv}(\bx,t)\equiv \int_\Omega d \by \ G^\Delta(|\bx-\by|) \bv(\by,t) \ , 
\ee
where $\Omega$ is the domain of definition of $\bv$, while for the sake of concreteness one can think  $G^\Delta$ as given by a projection 
operation in Fourier space, i.e. through spherically symmetric Galerkin truncation for all wavenumbers such that $k > 2 \pi/\Delta$.
\\

\noindent
In order to derive a hierarchy of equations relating the structure and correlation functions 
applicable to LES, we consider the filtered incompressible NSE on a three-dimensional
domain $\Omega = [0,L]^3$ with periodic boundary conditions    
\begin{align}
\label{eq:momentum-les}
&\p_t \ov_i + \p_j(\ov_i\ov_j + \overline{P}\delta_{ij} + \tau^\Delta_{ij}) = f_i \ , \\
&\p_i \ov_i  = 0 \ ,
\end{align}
where $P$ denotes the pressure, $\bm f$ the external force and 
$\tau^\Delta_{ij}=\overline{v_iv_j}-\ov_i \ov_j$ the SGS stress tensor, which 
is replaced by a model in LES applications. 
The density has been set to unity for convenience,
and the contribution of the viscous term is neglected.
The filter scale $\Delta$ is 
assumed to be smaller than the forcing scale $L_f$, such that $\overline{\bm f} = \bm f$. 
\\
\noindent
The aim of this paper is to study the exact equations that must be satisfied by the velocity structure functions, i.e. the  moments of the  resolved velocity-field increments
\be
\delta_\br \overline{\bv}(\bx) = \overline{\bv}(\bx+\br)-\overline{\bv}(\bx) \ .
\ee
The equation for  the second-order  correlation function, $\langle \delta_{\br} \overline{v}_i \delta_\br \overline{v}_j \rangle$,   has  been already derived 
in \cite{Meneveau94}. Here, we will further extend the previous results by generalising the exact hierarchy to  moments of any order  and by studying the relative importance of the different contributions entering in the corresponding equations of motion  by using {\em a-priori} and {\em a-posteriori} LES at high resolution.
The general evolution equation for the $n^{th}$-order correlation 
tensor consisting of velocity field differences is derived
from the momentum balance at points $\bx$ and $\bx'=\bx + \br$. Assuming homogeneity, we 
can  make a change of variables
$\bX  =  \frac{1}{2} (\bx + \bx') \ \mbox{ and } \ \br = \bx'- \bx \ ,$ and dropping all dependencies from $\bX$ in the averaged quantities. Furthermore, 
the partial derivatives 
with respect to $\bx'$- and $\bx$-coordinates will be  written as 
$
\partial_i' \equiv \partial_{x'_i} \ , 
\partial_i \equiv \partial_{x_i} \ ,
$
and homogeneity implies $ \p_i' = \p_{r_i} = \p_i$. 
Using the aforementioned results, one obtains the following evolution 
equation for the $n^{th}$-order correlation tensor  
for homogeneous isotropic turbulence
\begin{align}
\label{eq:les_increment_evol}
\p_t \langle \delta_r \ov_{i_1} \hdots \delta_r \ov_{i_n} \rangle  = &    
-\p_{k} \langle \delta_r \ov_{i_1} \hdots \delta_r \ov_{i_n} \delta_r \ov_k \rangle
\nonumber \\ 
& - \frac{1}{|S_{n-1}|}\sum_{\sigma \in S_n} \langle \delta_r \ov_{i_{\sigma(1)}} \hdots \delta_r \ov_{i_{\sigma(n-1)}} \delta_r (\p_k\overline{P}\delta_{k i_{\sigma(n)}}) \rangle
\nonumber \\
& + \frac{1}{|S_{n-1}|}\sum_{\sigma \in S_n} \langle \delta_r \ov_{i_{\sigma(1)}} \hdots \delta_r \ov_{i_{\sigma(n-1)}} \delta_r f_{i_{\sigma(n)}} \rangle   \nonumber \\
& - \frac{1}{|S_{n-1}|}\sum_{\sigma \in S_n} \langle \delta_r \ov_{i_{\sigma(1)}} \hdots \delta_r \ov_{i_{\sigma(n-1)}} \delta_r (\p_k\tau^\Delta_{k i_{\sigma(n)}}) \rangle \ ,
\end{align}
where $S_n$ denotes the symmetric group in $n$ elements, that is, we sum over 
all permutations of the indices $i_1, \hdots, i_n$.
In order to avoid counting identical terms involving 
products of $n-1$ velocity field increments multiple times, it is necessary 
to divide the sum over all permutations in $S_n$ by the number of elements 
of $S_{n-1}$ denoted by $|S_{n-1}|$.\\
It is important to stress that a similar  hierarchy of equations for the structure functions corresponding to the full 
Navier-Stokes evolution was derived in two different ways in  \cite{Yakhot01} and in \cite{Hill01}.
For comparison, the first two lines 
of eq.~\eqref{eq:les_increment_evol} are identical to eq.~(3.1) in Ref.~\cite{Hill01} obtained  for the
Navier-Stokes evolution, except for the absence of viscous terms in our case. 
The additional terms present in eq.~\eqref{eq:les_increment_evol} describe the effect of
the forcing in the third line and the
correlations between the velocity-field increments and the SGS tensor in the last line. 
This last term is the core object in the present paper, and our aim is study its 
scaling properties and its role in the balance equations. 
In the evolution equation for correlation tensors 
derived from the original NSE, 
the correlation with the viscous stress appears with the same structure as the correlation with the SGS-stress in Eq.~\eqref{eq:les_increment_evol}. 
\noindent
The hierarchy of equations in~\cite{Hill01} was derived from 
{\em kinematic constraints only}. These are: the  geometric constraints, i.e. restrictions on the form of the 
    correlation tensors due to their invariance under rotations and   reflections, and  incompressibility. Since no dynamical information was used in the derivation
of the  hierarchy for the full Navier-Stokes evolution, 
the algebraic {\em structure} of the LES hierarchy is  exactly the same. 
Furthermore, the additional correlation tensor in eq.~\eqref{eq:les_increment_evol}
must also obey the same kinematic constraints as the velocity increment tensors.  
Hence, in order to derive the LES hierarchy relating 
structure functions of any order, 
the only necessary work lies in the evaluation of the correlation tensors 
involving the SGS-stress, 
\be
H_{i_{1} \hdots i_{n}} \equiv \frac{1}{|S_{n-1}|}\sum_{\sigma \in S_n} \langle \delta_r \ov_{i_{\sigma(1)}} \hdots \delta_r \ov_{i_{\sigma(n-1)}} \delta_r (\p_k\tau^\Delta_{k i_{\sigma(n)}}) \rangle  \ .
\ee
The necessary calculations are summarised in Appendix \ref{app:derivation}.
Let us first introduce some general notations for the correlation functions
that will be met during the calculations. By restricting our
analysis to homogeneous and isotropic turbulence we can characterise all velocity correlation functions in terms of longitudinal and
$\delta_r v_L = (\overline{\bv}(\bx+\br)-\overline{\bv}(\bx))\cdot \br/r$ and the transverse,  $\delta_r v_N$, components, where the latter is any component of the vector 
$\delta_r \overline{\bv}_N = \delta_r \overline{\bv} - \delta_r v_L \br/r $. We will denote the correlation function made of $m$ longitudinal
and $n$ transverse velocity increments at scale $r$ as:
  \begin{align}
    D^{n,m}(r) \equiv \langle (\delta_r \ov_L)^n (\delta_r \ov_N)^m\rangle \ ,
    \label{eq:D}
    \end{align}
  and the multi-scale correlation functions including also the components of  the SGS stress tensor as:
  \begin{align}
    \label{eq:G}
    G^{n,m}_{i,j}(r,\Delta) \equiv \langle (\delta_r \ov_L)^n (\delta_r \ov_N)^m \tau^\Delta_{ij}\rangle \ . 
    \end{align}
It will be useful to introduce also  two more quantities for  correlation functions similar to Eq.~\eqref{eq:G}, namely:
\begin{align}
\label{eq:stensors-long}
S^n_{i,j}(r,\Delta) = \langle (\delta_r \ov_{L})^{n}  \tau^\Delta_{ki}\partial_{k}\ov_j  \rangle \ ,
\end{align}
and
\begin{align}
\label{eq:ttensors-long}
T^n_{i,j}(r,\Delta) = \langle (\delta_r \ov_{L})^{n}  \tau^\Delta_{ki}\partial_{k}'\ov_j'  \rangle \ ,
\end{align}
where for the last two cases for the sake of simplicity we have introduced
only the longitudinal velocity increments (for the set of exact
equations we are going to analyse in this paper it turns out that this choice
is not restrictive). 
Finally, the pressure correlations are denoted as
\begin{align}
\label{eq:ptensors}
P^n(r,\Delta) = \langle (\delta_r \ov_{L})^{n} \delta_r(\partial_{i}\overline{P}\delta_{iL})\  \rangle \ .
\end{align}
Let us note that we have retained the dependencies  on $r$ and $\Delta$ as appropriate, in order to stress the 
explicit dependencies  on the sub-grid stress tensor  and on its characteristic scale $\Delta$ when relevant. 
After some algebra, one derives the exact 
hierarchy obtained for the evolution of the general longitudinal 
$n^{\rm th}$ order structure function, $\partial_t D^{n,0}(r)$ 
(see Appendix \ref{app:derivation} and also \cite{Hill01,Yakhot01} for similar derivation obtained for the case of NSE):
\begin{align}
\label{eq:hierarchy}
\partial_t D^{n,0}(r) =& - \left( \p_r  D^{n+1,0}(r) + \frac{2}{r} D^{n+1,0}(r) - \frac{2n}{r} D^{n-1,2}(r) \right) \nonumber \\
& - 2n\left(\p_r +\frac{2}{r}  \right) G^{n-1,0}_{L,L}(r,\Delta)  + \frac{4n}{r}(G^{n-2,1}_{L,N}(r,\Delta)+ G^{n-1,0}_{N,N}(r,\Delta))\nonumber \\
& +  \frac{2n!}{(n-2)!}(S^{n-2}_{L,L}(r,\Delta)  + T^{n-2}_{L,L}(r,\Delta)) - 2nP^n(r,\Delta)  + 2nF^n(r)\ , 
\end{align}
where the correlations involving the forcing $F^n(r)$ are described in Appendix
\ref{app:derivation}. {The form and properties of the correlation
tensors are discussed in detail in Appendix \ref{app:gtensors} for 
those leading to the functions $G^{n-1,0}_{L,L}(r,\Delta)$, $G^{n-1,0}_{N,N}(r,\Delta)$ and,
$G^{n-2,1}_{L,N}(r,\Delta)$. The functions $T^{n-2}_{L,L}(r,\Delta)$ and $S^{n-2}_{L,L}(r,\Delta)$ including
their common combinatorial prefactor are treated in Appendix 
\ref{app:stensors}, and the pressure correlations are contained
 in Appendix \ref{app:ptensors}.} 
A similar set of equations can also be obtained for the evolution
of the most general mixed longitudinal-transverse case, $D^{n,m}(r) $. \\
Let us notice that not all terms are always present, as one can explicitly see
by rewriting the above relation for the first low order moments, $n=2$
(corresponding to the Monin-K\'arm\'an-Howarth energy balance equation), 
$n=3$ and $n=4$:  
\begin{align}
\label{eq:n2}
\p_t D^{2,0}(r)  =& -\left(\p_r+\frac{4}{r}\right) \left(\frac{1}{3}D^{3,0}(r) + G^{1,0}_{L,L}(r,\Delta) \right)  
                    + 4 S^{0}_{L,L}(\Delta) + 4F^2 \ , \\ \nonumber \\
\label{eq:n3}
\p_t D^{3,0}(r)  =& -  \left(\p_r+\frac{2}{r}\right) D^{4,0}(r) + \frac{6}{r} D^{2,2}(r) 
                    - 6\left(\p_r+\frac{2}{r}\right) G^{2,0}_{L,L}(r,\Delta) \nonumber \\
                &   + \frac{12}{r}(G^{1,1}_{L,N}(r,\Delta)+ G^{2,0}_{N,N}(r,\Delta)) 
                    + 12(S^{1}_{L,L}(r,\Delta)  + T^{1}_{L,L}(r,\Delta)) \nonumber \\
                &   - 6P^3(r,\Delta)  + 6F^3(r)\ , \\ \nonumber \\
\label{eq:n4}
\p_t D^{4,0}(r)  =& -  \left(\p_r+\frac{2}{r}\right) D^{5,0}(r) + \frac{8}{r} D^{3,2}(r) 
                    - 8\left(\p_r+\frac{2}{r}\right) G^{3,0}_{L,L}(r,\Delta) \nonumber \\
                &   + \frac{16}{r}(G^{2,1}_{L,N}(r,\Delta)+ G^{3,0}_{N,N}(r,\Delta))
                   +  24(S^{2}_{L,L}(r,\Delta)  + T^{2}_{L,L}(r,\Delta)) \nonumber \\
                &   - 8P^4(r,\Delta)  + 8F^4(r)\ . 
\end{align}

Equation \eqref{eq:n2} stands out from the hierarchy as the terms $P^2(r,\Delta)$, $T^2_{L,L}(r,\Delta)$,
$G^{1,0}_{N,N}(r,\Delta)$, $G^{0,1}_{L,N}(r,\Delta)$ and $D^{1,2}(r)$ are not present and the incompressibility
constraint  implies $P^2(r,\Delta)=0$ and $2D^{1,2}(r)/r = (\p_r +2/r)D^{3,0}(r)$.
 $T^2(r,\Delta)$, $G^{1,0}_{N,N}(r,\Delta)$ and $G^{0,1}_{L,N}(r,\Delta)$ can be absorbed into the derivative of $G^{1,0}_{L,L}(r,\Delta)$; see Appendices {\ref{app:fourfifth} and 
 \ref{app:stensors}} for further details.
It is important to notice that the function $S^{n-2}_{L,L}(r,\Delta)$ for $n=2$ in
Eq.~\eqref{eq:n2} is not a function of $r$, i.e. it is  not a real multi-scale function, since 
\be
 S^{0}_{L,L}(\Delta) = \langle \tau^\Delta_{kL}\p_k\tau_{kL} \rangle \
\ee
 is proportional to the SGS energy transfer: 
$$3S^{0}_{L,L}(\Delta) = \Pi(\Delta) = -\langle \tau^\Delta_{ki}\p_k \ov_i \rangle,$$  see { Appendices \ref{app:fourfifth} and \ref{app:stensors}} or Ref.~\cite{Meneveau94}, where Eq.~\eqref{eq:n2} has been derived directly. 

\noindent
Before proceeding, let us make a few general comments about the structure of
the different terms entering in Eq.~\eqref{eq:hierarchy}. It is
important to notice that the terms containing correlations of type $
G^{\alpha,\beta}_{i,j}(r,\Delta)/r$ with $(\alpha,\beta) = (n-1,0), \, (n-2,1)$
and the term $S^{n-2}_{i,j}(r,\Delta)$ have the same physical dimensions
but two completely different roles: the former consists of $n-1$ velocity-field increments multiplied by the SGS stress tensor, the latter consists of $n-1$ velocity-field increments multiplied by terms of the form  $\tau^\Delta_{ki} \partial_k \ov_j$ that
contribute to the definition of  the SGS energy transfer, $\Pi(\Delta)$. 
As a result, the latter will play a key role in the balancing of
the hierarchy  as suggested from the fact that also in the
original NSE the presence of the dissipative anomaly is a signature of
non-trivial multi-scale correlation functions among viscous  and inertial
scales. On the contrary, the terms labelled $T^{n-2}_{i,j}(r,\Delta)$ are not correlated to the local energy transfer being defined in terms 
of the SGS stress tensor and the velocity gradient  at two different
points $ \bx$ and $\bx'$.

\section{Scaling of correlation functions}
\label{sec:scaling}

In this section, we will first assess the scaling properties of 
all terms entering  in the previous  hierarchy 
(\ref{eq:hierarchy}) from a phenomenological point of view. 
In Section \ref{sec:numerics}, we will check using  DNS and LES
what is observed in reality and whether SGS modelling based on
the Smagorinsky eddy viscosity is indeed able to reproduce the correct
observations.  \\ 
 A popular and
fruitful way to phenomenologically introduce intermittency in turbulence theory
is to suppose that the velocity field is described by a MF
process, where the velocity increment scales with a local H\"older exponent 
$h$, that is, $\delta_r v \sim r^h$, on a fractal set of dimension $ D(h)$. Such
phenomenological hypothesis has been used in the past to explain the observed 
anomalous scaling properties of the single-scale longitudinal and
transverse velocity structure functions, 
the distribution of velocity gradients, of particles' accelerations, 
velocity increments along particle trajectories and
many other single and multi-scale turbulent properties 
\cite{bec2006acceleration,hill2002scaling,arneodo2008universal,
benzi2010inertial,biferale2004multifractal,la2001fluid}.
The simplest way to build up a MF-signal is to embed the velocity field into a
multiplicative process, supposing that the velocity-field fluctuations at 
two nested,
inertial-range, scales $ r_1 < r_0$ are connected by a scaling relation:
  \be
  \delta_{r_1} v = \left(\frac{r_1}{r_0}\right)^h \delta_{r_0} v 
  \ee
and imagining that the successive breaking into eddies at smaller scale $r_2
<r_1$ will be given by another multiplicative process with a different, but
identically distributed, realisation of the local exponent, $h'$ 
\cite{benzi1984multifractal,she1994universal,dubrulle1994intermittency,meneveau1987simple}
\be
\delta_{r_2} v = \left(\frac{r_2}{r_1}\right)^{h'} \delta_{r_1} v.
\ee
Using this approach it is possible to predict the scaling behaviour for all
terms entering in the hierarchy (\ref{eq:hierarchy}).  We examine now the most
important ones.

\subsection{Single-scale Structure Functions $D^{n,m}(r)$}
From the multiplicative MF Ansatz and by assuming that longitudinal and
transverse increments do follow the same scaling distribution, it is 
straightforward
to predict that \cite{Frisch91} 
\be
\label{eq:Dmf}
D^{n,m}(r) \sim \int dh \left ( \frac{r}{L_0} \right )^{h(m+n)} \left ( \frac{r}{L_0} \right )^{3-D(h)} (\delta_{L_0} v)^{n+m} \sim A_{n,m} \left ( \frac{r}{L_0} \right )^{\zeta_{n+m}}  
\ee
where the last equality is obtained by estimating the integral in the saddle
node approximation, $ r \ll L_0 \to \zeta(n+m) = \min_h(h(m+n) + 3-D(h))$. 
The prefactors $A_{n,m}=O(1)$ are  non-universal 
quantities which
depend on the large-scale velocity distribution $\delta_{L_0} v$. 
One can immediately see
that as soon as multiple realisations of the
local H\"older exponent exist, 
the scaling properties are characterised by anomalous
power laws, i.e. $\zeta(n) \neq n/3$. Nevertheless, it is important to notice
that the MF approach contains the Kolmogorov K41 phenomenology as a limiting 
case, where the energy cascade is assumed to develop in a homogeneous way 
with a H\"older-$1/3$ velocity field everywhere in the three-dimensional 
volume, since $h= 1/3$ and $D(1/3)=3$ imply $\zeta(n) = n/3$. 

\subsection{Multi-scale  correlation functions and \textbf{\textit{Fusion-Rules}} } 
Using the same approach,  one can 
show that multi-scale correlation functions
must also be characterised by anomalous scaling properties. For the generic
two-scale correlation functions with $r <R$ we have the  {\it
Fusion-Rules} (FR) behaviour \cite{Lvov96a,Lvov96b,Benzi98}:

\begin{align}
\label{eq:FR}
& \langle (\delta_r v )^n  (\delta_R v)^m \rangle  \sim  \langle \left (\frac{r}{R}\right )^n (\delta_R v)^n  (\delta_R v)^m \rangle  \sim \nonumber \\
& \sim \int dh dh' \left (\frac{r}{R}\right )^{hn+3-D(h)} \left (\frac{R}{L_0}\right )^{h'(n+m)+3-D(h')} \nonumber 
\\ 
& \sim \left(\frac{r}{R}\right)^{\zeta_n} \left(\frac{R}{L_0}\right)^{\zeta_{n+m}}; \qquad r \ll R \ll L_0
\end{align}
where we have assumed a large separation among all scales, that $r$ and $R$
belong to the inertial range and we have applied a double saddle-node
approximation of the integrals.  
Notice that Eq.~(\ref{eq:FR}) would correspond to the uncorrelated
result {\it iif} the exponent follows K41, $ \zeta_n = n/3 $,
  \be
   \langle (\delta_r v )^n  (\delta_R v)^m \rangle  \sim r^{n/3} R^{m/3}.
  \ee

\subsection{Multi-scale Correlation among velocity increments and SGS-stress, $G^{n,m}_{i,j}(r,\Delta)$ }
In order to introduce multi-scale correlation with the SGS stress tensor and
the SGS energy dissipation entering in the hierarchy (\ref{eq:hierarchy}) we
start from the observation made by \cite{Constantin94,eyink2,Vreman94a} that
the local SGS stress tensor can be estimated in terms of a suitable average of
local velocity increments. As a result, for any H\"older-continuous velocity
fields with  local H\"older exponent $h$ one might estimate $\tau^\Delta$ to be
a (local) MF-scaling function of the coarse-graining grid
$\Delta$ \cite{Eyink96a,Aluie2009I,Aluie2009II,Buzzicotti17a}:
\be
\tau^\Delta_{ij} \sim \Delta^{2h}
\ee
Any correlation tensor involving velocity field increments at scale $r$ and the
SGS-stress $\tau^\Delta_{ij}$ can therefore be treated as a correlation tensor
involving the two scales $r$ and $\Delta$ within the {\it Fusion-Rules}
approach.  
Following the same MF  Ansatz of the previous section, we end up with:
\be 
\delta_\Delta \ov_L \sim \left (\frac{\Delta}{r}\right )^h \delta_r \ov_L \ .
\ee
For $\Delta < r$, the scaling behaviour of the correlations between the $n^{\rm
th}$ power of a longitudinal velocity field increment and the SGS-stress can be
estimated  using Eq.~(\ref{eq:FR}): 
\be
\label{eq:FRSGS}
\langle  \tau^{\Delta}_{LL} (\delta_r\ov_L)^n\rangle 
\sim \left \langle  (\delta_\Delta v_L)^2  (\delta_r \ov_L)^n  \right \rangle  
\sim \left \langle \frac{\Delta^{2h}}{r^{2h}} (\delta_r \ov_L)^{n+2} \right \rangle
\sim \left (\frac{\Delta}{r} \right)^{\zeta_2} \left (\frac{r}{L} \right)^{\zeta_{n+2}} \ ,
\ee 
hence
\be
\label{eq:mf-scaling-estimate}
\langle (\delta_r\ov_L)^n \tau^{\Delta}_{LL}\rangle \sim 
\left (\frac{\Delta}{L} \right)^{\zeta_2} \left (\frac{r}{L} \right)^{\zeta_{n+2}-\zeta_2} \ . 
\ee
A few comments are now in order. First, the FR approach, being based on a MF
multiplicative cascade, does not easily incorporate differences among scaling
properties of longitudinal or transverse velocity increments. In fact,  the
most recent literature \cite{iyer2017reynolds} shows that  such a differences might
disappear with increasing Reynolds numbers. Hereafter we will always assume that it is not important to 
distinguish among  scaling properties of longitudinal, transverse or mixed
longitudinal-transverse components, i.e.  in all cases only the total number of
velocity increment matters. Second, the FR estimate (\ref{eq:FRSGS}) is meant
to capture only the leading power law behaviour and cannot take into account cancellations and symmetry constraints which may affect the prefactors.
For example, the prediction (\ref{eq:FR}) cannot hold for the special case of
mixed longitudinal-transverse correlation with an odd power for the transverse
increment, because in such a case $D^{n,2m+1}(r) =0, \, \forall m$ because of 
isotropy \cite{Monin75}. We will come back to this point in
Sec.~\ref{sec:numerics} where we analyse the data
from DNS and LES.  
For the sake of comparison, it will  be important to  estimate the  multi-scale
correlation functions by assuming  that the fields at different scales are
almost decorrelated: 
\begin{align}
\label{eq:uc-scaling-estimate}
\langle (\delta_r\ov_L)^n \tau^{\Delta}_{LL}\rangle 
\sim \left \langle  \Big(\frac{r}{L}\Big)^{nh} \Big(\frac{\Delta}{L}\Big)^{2h'}  \right  \rangle 
\sim \left(\frac{r}{L}\right)^{\zeta_n} \left(\frac{\Delta}{L}\right)^{\zeta_2} \ .
\end{align}
Since $\zeta_{n+2}-\zeta_2 < \zeta_n$ and $r/L<1$, the uncorrelated
scaling Ansatz would be subleading with respect to that obtained from the MF
cascade process.  We will return to this point in Sec.~\ref{sec:numerics}. 

\subsection{Correlations between velocity field increments and components of the  SGS-energy, $S^n_{i,j}(r,\Delta)$ }
The multiplicative cascade Ansatz can also be used to estimate the scaling
behaviour of the correlations functions involving the components of the
SGS-energy transfer
\be
\label{eq:S}
S^n_{i,j} = \langle \tau^\Delta_{ki}\partial_{k}\ov_j(\delta_r \ov_L)^n \rangle.
\ee
As already noticed in the previous subsection, we will assume that no major
scaling differences exist concerning the longitudinal or 
the transverse components of the different observables,  
and we proceed by applying the MF approach by specifying it for the case where 
all components are chosen in the longitudinal directions. 
\begin{align}
\langle \tau^\Delta_{LL}\partial_{x_L}\ov_L(\delta_r \ov_L)^n \rangle 
& \sim \left \langle (\delta_\Delta v_L)^2\frac{\delta_\Delta v_L}{\Delta} (\delta_r \ov_L)^n \right \rangle \sim 
\\ \nonumber
& \sim \left \langle \frac{\Delta^{3h-1}}{r^{3h}} (\delta_r \ov_L)^{n+3} \right \rangle
\sim 
\frac{1}{\Delta}\left (\frac{\Delta}{r} \right)^{\zeta_3} \left (\frac{r}{L} \right)^{\zeta_{n+3}} \ .
\end{align}
Using the exact scaling property $\zeta_3 = 1$ one obtains
\be
\label{eq:pi-scaling-estimate-casc}
\langle \tau^\Delta_{kL}\partial_{k}\ov_L (\delta_r \ov_L)^n \rangle 
\sim r^{-1} \langle (\delta_r \ov_L)^{n+3} \rangle \sim r^{\zeta_{n+3}-1}
\ee
hence, in the inertial-range scaling regime, all curves
obtained at different $\Delta$ must collapse.  Before summarising all  results,
let us mention that the scaling of the pressure terms, $P^n(r,\Delta)$ in Equation
(\ref{eq:hierarchy}) will necessarily be connected to a mixing of all previous
correlation functions, because it feels contributions from 
both the
advection term $\ov_i \partial_i \ov_j$ and SGS tensor in Equation
(\ref{eq:momentum-les}). On the contrary, one  expects that the terms
involving $T^n_{i,j}(r,\Delta)$ will always be sub-leading with respect to
$S^n_{ij}(r,\Delta)$, because it consists of velocity gradients and SGS stress
components in two different spatial locations. \\

\noindent More importantly, the above scaling relations tell us that the contribution
involving the correlation with the components of the SGS energy transfer in Equation
(\ref{eq:hierarchy}) are independent of $\Delta$ and they have the same
contributions as the non-linear single-scale structure-function terms
\begin{equation} 
\partial_r D^{n+1,0}(r) \sim r^{-1}D^{n+1,0}(r) \sim r^{-1} D^{n-1,2}(r) \sim
S^{n-2}_{L,L}(r) \sim r^{\zeta_{n+1}-1} \ ,
\label{eq:MFleading}
\end{equation}
while the terms involving correlations
with the SGS stress are subleading in the limit $\Delta/r \to 0$, 
and do depend on the cut-off $\Delta$ 
\begin{equation}
r^{-1} G_{L,L}^{n-1,0}(r,\Delta)  \sim r^{-1} G_{L,N}^{n-2,1}(r,\Delta) \sim  r^{-1} G_{N,N}^{n-1,0}(r,\Delta) \sim \left (\frac{\Delta}{r}\right )^{\zeta_2} r^{\zeta_{n+1}-1} \ . 
\label{eq:MFsubleading}
\end{equation}

\section{Numerical results}
\label{sec:numerics}
\begin{figure}[H]
\includegraphics[width=\textwidth]{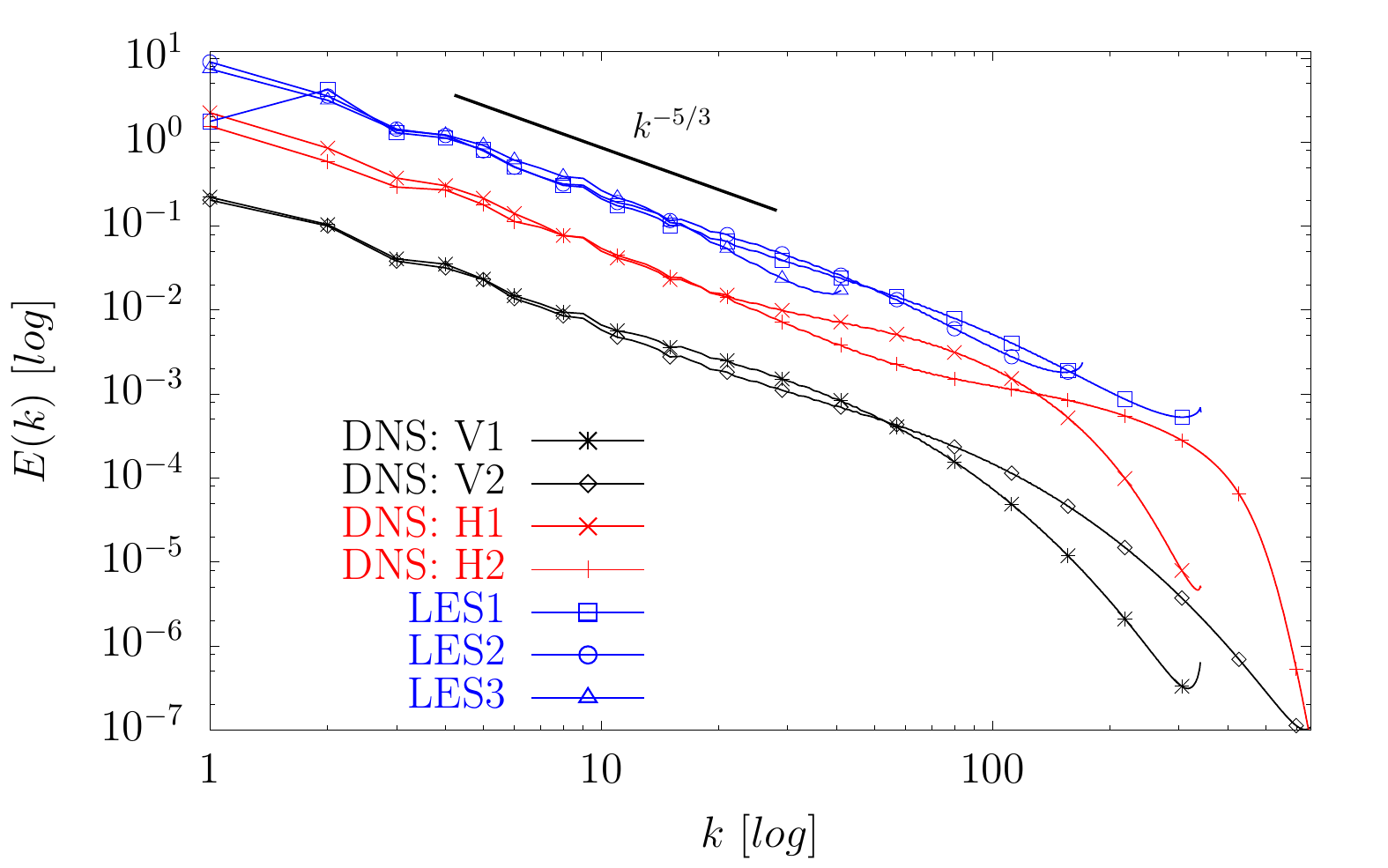}
\caption{
Energy spectra, $E(k)$, averaged on time in the stationary state for the
different sets of simulations. The spectra for the LES are shifted upward while
the spectra for data-sets with normal viscosity (V) are shifted downward.
}
\label{fig:simulations_spectra}
\end{figure}

In order to measure scaling exponent, to compare them to the derived scaling 
results, and to establish which terms in the balance equations are leading or sub-leading, we need to generate data-sets for both 
{\em a-priori} and {\em a-posteriori} analyses.  For the
{\em a-priori} analysis, data-sets are generated through DNSs of the
viscous  and hyper-viscous NSE  
\begin{align}
\label{eq:nse}
&\partial_t \bv = - \nabla \cdot (\bv \otimes \bv) - \nabla p + \nu (-1)^{\alpha + 1}\Delta^{\alpha}\bv + \bF \ , \\
\label{eq:incomp}
&\nabla \cdot \bv = 0 \ ,
\end{align}
where $\bv$ denotes the velocity field, $p$ the pressure divided by the
density, $\bF$ an external force, $\alpha$ the power of the Laplacian and
$\nu$ the kinematic (hyper)viscosity. We carry out series of numerical 
simulations with either normal viscosity ($\alpha =1$) or hyperviscosity 
($\alpha =2$ and $\alpha =4$), the data-sets are distinguished by the labels
V (visco) and H (hyperviscous), respectively.
The DNS velocity fields are subsequently filtered    
through spherically symmetric Galerkin truncation at a cut-off wavenumber 
$k_c = \pi/\Delta$ \cite{Pope00}, i.e. $G^\Delta$ is given by a projection
operation in Fourier space.
For the {\em a-posteriori} analysis, LESs are carried out following 
Eq.~\eqref{eq:momentum-les} using the standard static Smagorinsky model 
for the deviatoric part of the SGS stress tensor 
\begin{align}
\label{eq:smagorinsky}
\tau^{\Delta,\rm SMAG}_{ij} = -2 (c_s^\Delta \Delta)^2 \sqrt{\os_{ij}\os_{ij}} \, \os_{ij} \, { = -2 \nu_{_E}\os_{ij}} \ ,
\end{align}
{where $\nu_{_E}=(c_s^\Delta \Delta)^2 \sqrt{\os_{ij}\os_{ij}}$ is the scalar eddy viscosity}, $c_s^\Delta$ is the Smagorinsky constant which is here set to $c_s^\Delta=0.16$ and $\os_{ij} = 1/2 (\partial_j \ov_i+\partial_i \ov_j)$ is the resolved strain-rate tensor \cite{Lilly67,MeneveauARFM}. The respective evolution
equations for DNS and LES are solved numerically on a domain $\Omega =
[0,2\pi]^3$ with periodic boundary conditions using the pseudospectral method
with full dealiasing according to the $2/3$rds rule \cite{Patterson71}.  In
both cases the large-scale forcing was given in Fourier space by a second-order
Ornstein-Uhlenbeck process, which is active in the wavenumber band $k \in
[0.5,1.5]$ \cite{Sawford91,BiferalePRX2016}, 
{corresponding to the forcing scale $L_f = 2 \pi/k_f =4.2$, where $k_f =1.5$ is the upper limit of the forcing interval.}. The
resolution for the DNSs is $\eta_\alpha/dx \simeq 0.7$ for all simulations,
where $dx$ is the grid spacing and $\eta_\alpha = (\nu^3/\epsilon)^{1/6\alpha
-2}$ the generalised Kolmogorov microscale \cite{Borue95} with $\varepsilon$ denoting
the mean dissipation rate.  
After reaching a statistically stationary state the DNS and LES 
velocity fields and the LES SGS-tensor have been
sampled at intervals of one large-eddy turnover time in order to create 
ensembles of statistically independent data, from which all correlation 
functions are calculated. 
{Concerning the resolution of the DNSs, runs V1 and H1 are carried out on $1024^3$ collocation points while 
$2048^3$ collocation points were used for runs V2 and H2. For LES, 
grids of size $128^3$, $512^3$ and $1024^3$ we used, the corresponding 
runs are labelled LES1, LES2 and LES2.
}
Further details of all
DNS and LES 
are given in table \ref{tbl:simulations}.
Steady-state energy spectra of all data-sets are shown in 
Fig.~\ref{fig:simulations_spectra}.

\begin{table}[H]
 \begin{center}
\begin{tabular}{cccccccccc}
   Data & $N$ & $\RL$ & $\varepsilon, \max[\Pi]$ & $U_0$ & $L_0$ & $\nu$ & $\alpha$ & $\frac{T_0}{T_{\rm eddy}}$ & $\Delta$ \\ 
  \hline
V1 & 1024 & 2570 & 1.9 & 1.8 & 1.2 & 0.0008 & 1 & 25 & $\pi/12-\pi/40$ \\
V2 & 2048 & 8000 & 1.4 & 1.5 & 1.2 & 0.0003 & 1 & 9 & $\pi/12-\pi/40$ \\
H1 & 1024 & 8000 & 1.9 & 1.9 & 1.3 & $2 \times 10^{-8}$ & 2 & 7 & $\pi/12-\pi/40$ \\ 
H2 & 2048 & 26000 & 1.5 & 1.6 & 1.1 & $5.7 \times 10^{-20}$ & 4 & 6 & $\pi/80$ \\ 
  \hline
LES1 & 128 & - & 1.3 & 1.5 & 1.2 & 0 & - & 190 & $\pi/41$ \\ 
LES2 & 512 & - & 1.5 & 1.7 & 1.3 & 0 & - & 10 & $\pi/171$ \\ 
LES3 & 1024 & - & 1.3 & 1.4 & 0.8 & 0 & - & 27 & $\pi/342$ \\ 
  \hline
  \end{tabular}
  \end{center}
 \caption{
The DNSs have been carried out with either normal or hyperviscosity,
where $\alpha$ is the order of the Laplacian and the corresponding data-sets
are identified by the labels V1, V2 and H1, H2, respectively.
$N$ denotes the number of grid points in each Cartesian coordinate,
$U_0$ the RMS velocity, $L_0=(\pi/2U_0^2)\int dk \ E(k)/k$ the integral scale,
$\nu$ the kinematic hyperviscosity, 
$\varepsilon$ the dissipation rate which equals the maximal 
inertial flux $\max[\Pi]$ in steady state, $T_0/T_{\rm eddy}$ the
steady-state run time in units of large-eddy turnover time $T_{\rm eddy}=L_0/U_0$, 
and $\Delta= \pi/k_c$ the filter scale in terms of the cut-off wave number $k_c$. 
The values given for $\varepsilon$, $U_0$, $L_0$ and $\max[\Pi]$ 
are time averages, where $\max[\Pi]$ is reported for LES while $\varepsilon$
 is reported for DNS.  
The integral-scale Reynolds number 
is defined as $\RL = C (L_0/l_d)^{4/3}$, where $C$ is a
constant estimated by comparison to data-set V1 and $l_d$ is
the scale corresponding to the maximum of $k^2 E(k)$; for further details see 
Ref.~\cite{Buzzicotti17a}. 
{The forcing scale $L_f = 2 \pi/k_f =4.2$, with $k_f =1.5$ being the maximum wavenumber where the forcing is applied, is the same for all simulations.} 
}
 \label{tbl:simulations}
\end{table}

\subsection{Second-order balance $(n = 2)$}
\begin{figure}[H]
\includegraphics[width=\textwidth]{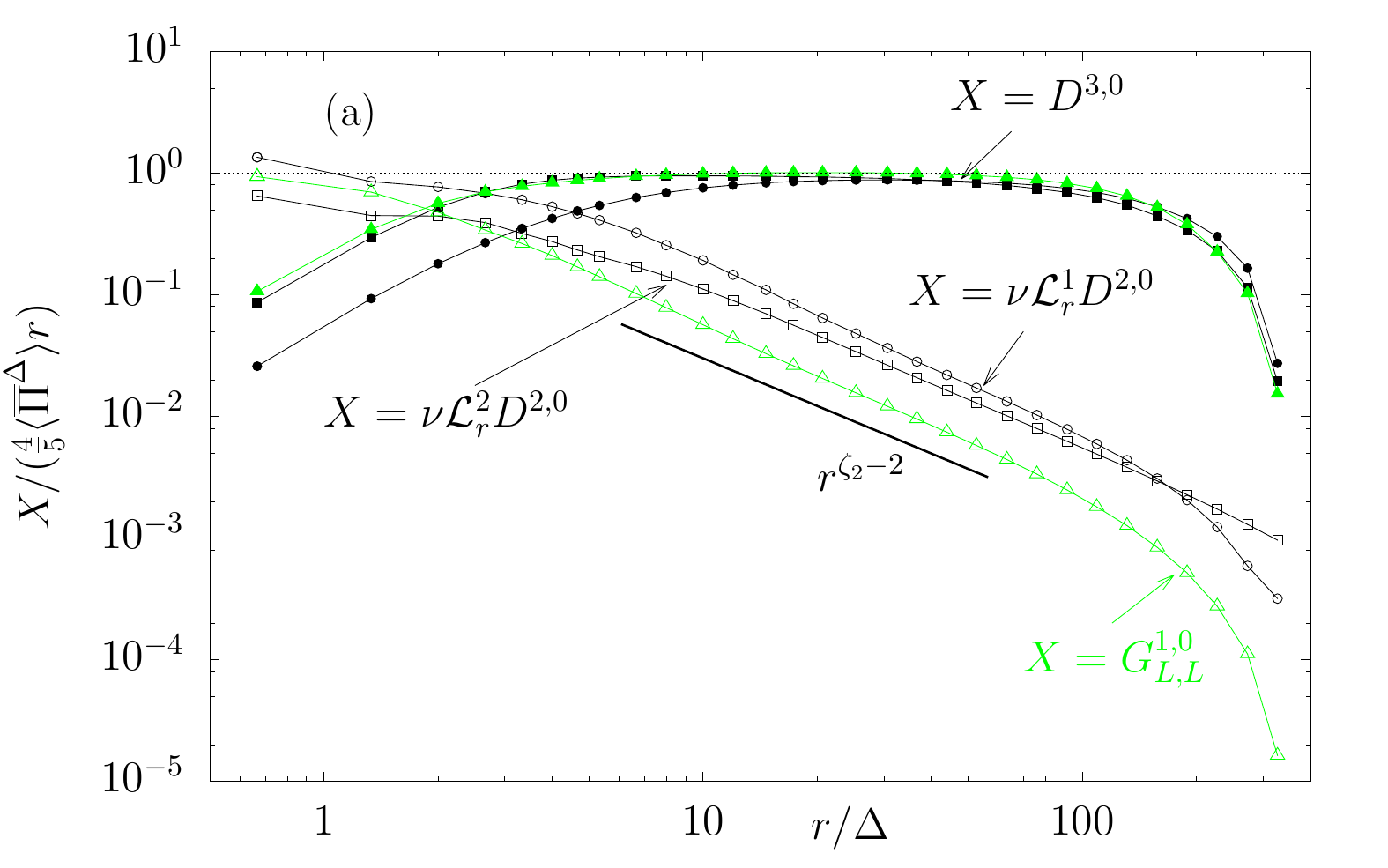} \\ 
\includegraphics[width=\textwidth]{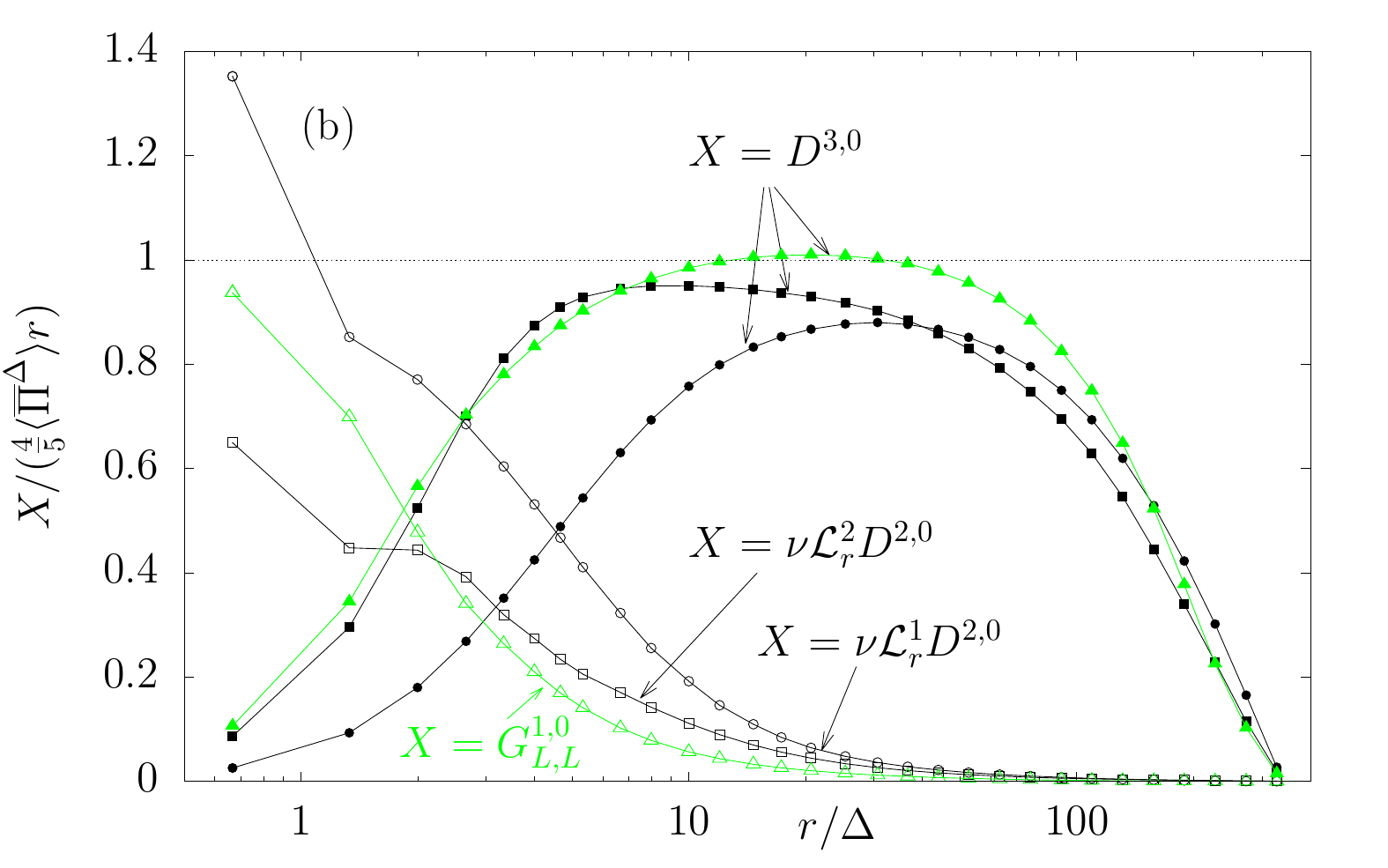}
\caption{
Four-fifth law: comparison between full DNSs' data with normal viscosity (V1, black circles), with hyperviscosity (H1) (black squares) and the Smagorinsky LES data at $\Delta = \pi/342$ (green/light grey triangles) always using $1024^3$ grid points. Full symbols are the third order structure functions $D^{3,0}(r)$ while open symbols represent either the viscous term,  $\nu \mathcal{L}^\alpha_r D^{2,0}_{L,L}(r)$ (black colour), or the correlation with the SGS tensor, $G^{1,0}(r,\Delta)$ (green/light grey). All terms are normalised with $4/5\langle \overline{\Pi}(\Delta) \rangle r$. 
Panel (a): data are presented in log-log scale to show the scaling properties of the different functions. Panel (b): same data in lin-log scale to highlight how the four-fifth law expected value (dashed line) is recovered by the different structure functions in the inertial range.
}
\label{fig:smag-fourfifth-fullDNS}
\end{figure}

Following Ref.~\cite{Meneveau94} and eq.~\eqref{eq:n2}  we obtain  the equivalent of the four-fifth law within the  LES formulation which reads in the stationary state:
\be
\frac{1}{r}\left(D^{3,0}(r) + 3 G^{1,0}_{L,L}(r,\Delta)\right) =  -\frac{4}{5} \langle \overline{\Pi}(\Delta) \rangle \ ,
\label{eq:45}
\ee 
where we have neglected the forcing contribution because it is always sub-leading for scales smaller than the forcing scale $L_f$; 
{see Appendix \ref{app:fourfifth}}. 
In Fig.~\ref{fig:smag-fourfifth-test}(a), we show the importance of the two terms on the left-hand side, using both  the filtered DNS data at 
$12 \leqslant k_c \leqslant 40$ and the data from the LES1 simulation. Panels (b) and (c) of the same figure show the same curves for LES data only upon changing the resolution.   
It is clear that the LES approach does not introduce any important spurious physics in the inertial range if compared either with the viscous or the hyperviscous simulations. In particular, panel (a) shows that the LES curves are recovered from the {\em a-priori} analysis by decreasing the filter cut-off. The solid lines $(r/\Delta)^{-\zeta_2}$ in
Figs.~\ref{fig:smag-fourfifth-test}(a-c) 
indicate the MF prediction for $G_{L,L}^{1,0}(r,\Delta)/r$, {which for $n=1$ and $m=0$ would give $G_{L,L}^{1,0}(r,\Delta)/r \sim r^{\zeta_3-\zeta_2-1}=r^{-\zeta_2}$}. It is  clear from the figures that $G_{L,L}^{1,0}(r,\Delta)$  
does not obey the MF scaling in both the {\em a-priori} and the  {\em a-posteriori} analyses. 
Instead, interestingly enough, it is  even more sub-leading than the MF prediction,  indicating that the details of the SGS-model should have little effect on the energy balance. The deviation from the MF in the DNS data is probably due to the existence of cancellations given  the particular structure of $G_{L,L}^{1,0}(r,\Delta)$ where the longitudinal increments appear only in a linear way, a fact that would lead to an exactly vanishing contribution in the case of weak correlation with the SGS stress tensor, because of homogeneity. For the LES case we will comment on this later on in this section.
Since 
$D^{3,0}(r) \sim r^3$ for $r \to 0$, $G^{1,0}_{L,L}(r,\Delta)$ must satisfy 
$-3G^{1,0}_{L,L}(r,\Delta)= \frac{4}{5} \langle \Pi(\Delta) \rangle$  for 
$r \to 0$. This is the case
as can be seen in Fig.~\ref{fig:smag-fourfifth-test}(a)-(c) for both the filtered DNS and the LES. 
Moreover, the scaling range of the correlation and structure 
functions obtained though the LES simulations extends with increasing resolution as shown in 
Figs.~\ref{fig:smag-fourfifth-test}(b,c), 
as it must be expected for a good subgrid parametrisation.  \\   Finally, we compare in Fig.~\ref{fig:smag-fourfifth-fullDNS} the highest resolved LES3 data-set against results from the viscous 
and hyperviscous data-sets V1 and H1 without filtering, where all simulations were carried out on
$1024^3$ grid points. Here, for the two data sets from the Navier-Stokes cases we need to consider that the $4/5$ law  (\ref{eq:45}) will  include the dissipative
term: {$\nu \mathcal{L}^\alpha_rD^{2,0}(r)$, where 
$\mathcal{L}^\alpha_r$ is a differential operator which depends on the order $\alpha$ of the Laplacian. The dissipative term replaces 
term $G^{1,0}$ in the $4/5$-th law; for data-set V1 it is given as 
$\mathcal{L}^1_r D^{2,0}(r)/r= 6\partial_r D^{2,0}(r)/r$
while for data-set H1 
$\mathcal{L}^2_r D^{2,0}(r)/r= 12 D^{2,0}(r)/r^2 + 12 (\int_0^r ds \ s D^{2,0}(r))/r^4  - 6 \partial_r^3 D^{2,0}(r)/r$.
}

For data-set V1, {the form of} the dissipative term {implies that it} should scale as 
$r^{\zeta_2-2} \approx r^{-1.3}$, which is well satisfied, as can be seen from panel (a). Interestingly, the function $G^{1,0}_{L,L}(\Delta,r)/r$ scales similarly as a function of  $r$ at fixed $\Delta$; {a possible explanation for this behaviour is given below in eq.\eqref{eq:smag}}. For data-set H1, 
$\mathcal{L}^2_r D^{2,0}(r)/r= 12 D^{2,0}(r)/r^2 + 12 (\int_0^r ds \ s D^{2,0}(r))/r^4  - 6 \partial_r^3 D^{2,0}(r)/r$, 
and therefore,
{at leading order 
$\mathcal{L}^2_r D^{2,0}(r)/r \sim D^{2,0}(r)/r^2$
for $r > \eta_{\alpha}$}, 
does show the same scaling properties as the viscous case. Panel (b) shows that the third-order structure function obtained from  LES3 has an inertial-range scaling much more extended than the viscous case and even better than the H1, supporting the statement that the LES closure is a dissipative closure more efficient than hyperviscosity. Overall, we can conclude that if the use of the hyperviscosity is interpreted as an effective `subgrid model',  it leads to a larger influence of the dissipative term than the SGS modelling of the LES simulation.  

\begin{figure}[H]
\centering
\includegraphics[width=\textwidth]{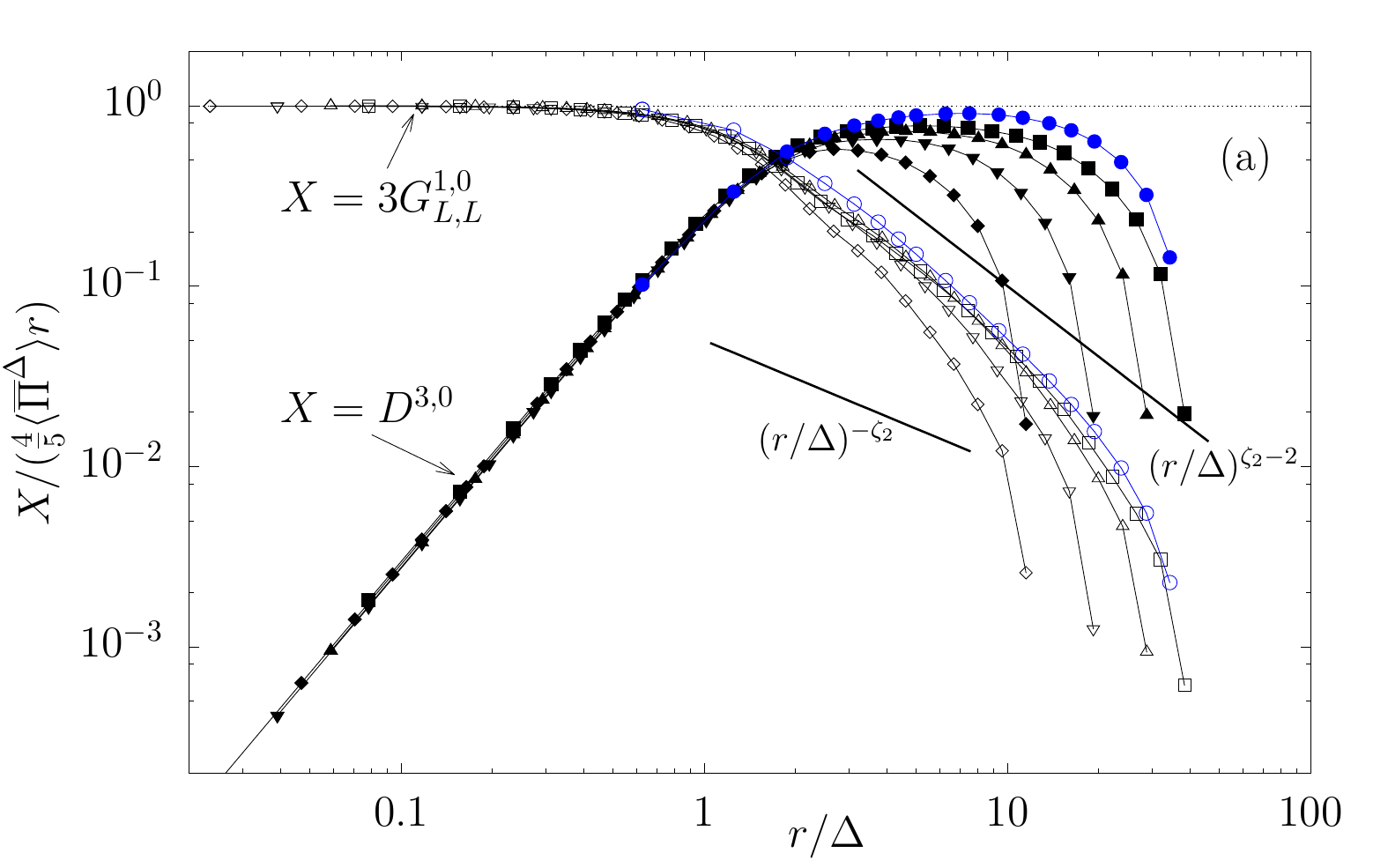} \\ 
\includegraphics[width=0.48\textwidth]{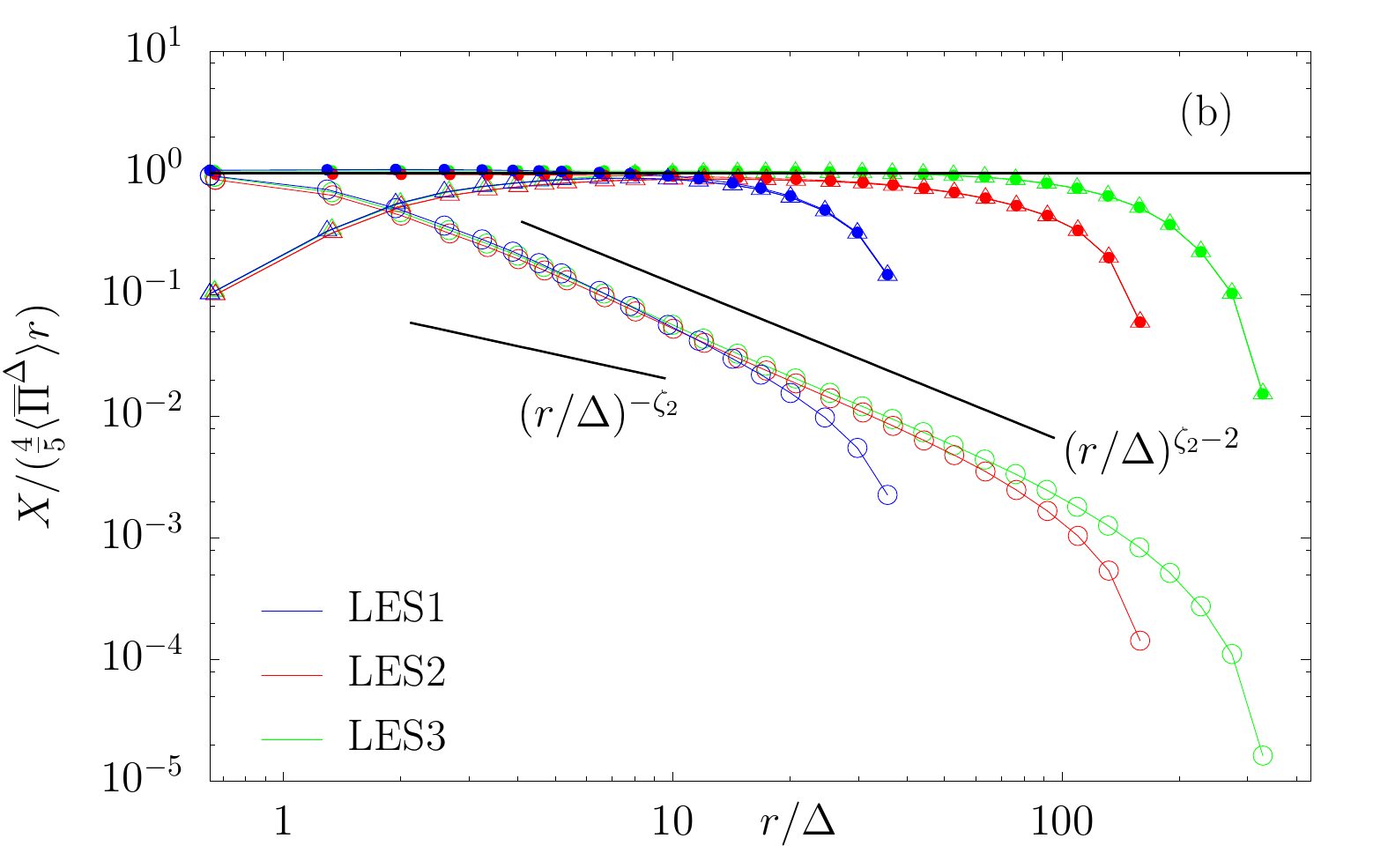}\hspace{1em}
\includegraphics[width=0.48\textwidth]{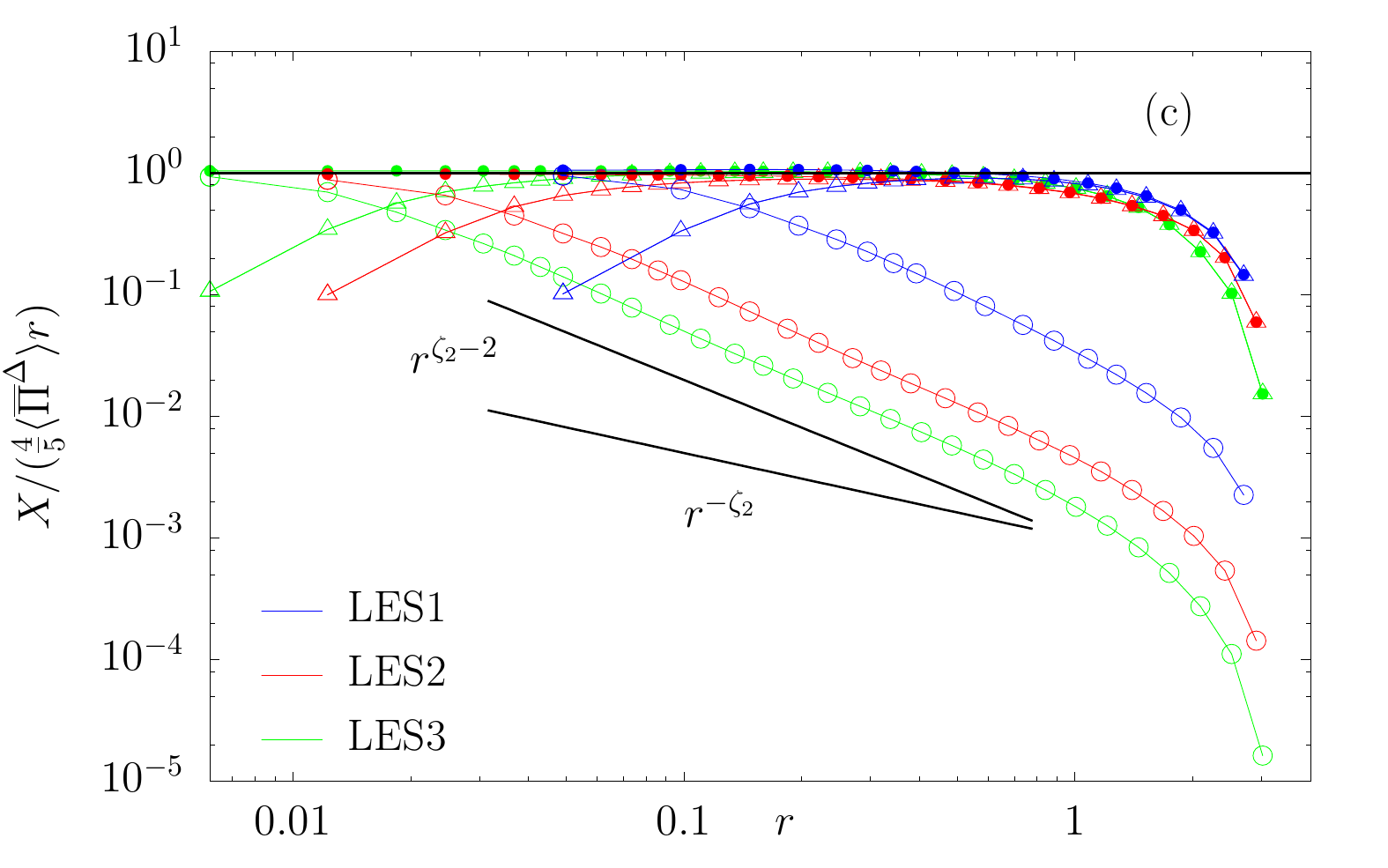}
\caption{ 
Top panel (a): comparison of the two terms in the left hand side of the four-fifth law, eq.~\eqref{eq:45}. Full symbols represent $D^{3,0}(r)$ while open symbols are $3 G^{1,0}(r,\Delta)$, both normalised with $4/5\langle \overline{\Pi}(\Delta) \rangle r$. The two terms are calculated for the LES1 data-set at $\Delta = \pi/42$ (circles) and for the filtered DNS data using different cut-off $\Delta$, (black rhombus) $\Delta = \pi/12$, (black downwards triangles) $\Delta = \pi/20$, (black upwards triangles) $\Delta = \pi/30$ and (black squares) $\Delta = \pi/40$.
Bottom panels (b and c): the same quantities, $D^{3,0}(r)$ (full triangles) and $3 G^{1,0}(r,\Delta)$ (open circles) are presented for different LES data-sets, namely: LES1, $\Delta = \pi/42$ (blue coluor/dark grey). LES2,  $\Delta = \pi/171$ (red/grey) and LES3 $\Delta = \pi/342$ (green/light grey) against $r/\Delta$ in panel (b) and against $r$ in panel (c). {The solid line $(r/\Delta)^{-\zeta_2}$ shown in all panels, indicates the MF scaling prediction \eqref{eq:mf-scaling-estimate}, while the solid line $(r/\Delta)^{\zeta_2-2}$ corresponds to the scaling prediction based on the assumption of a constant eddy viscosity \eqref{eq:smag}.}

}
\label{fig:smag-fourfifth-test}
\end{figure}

Before moving to the balance equation of higher-order correlations,  let us have a look in more details at the scaling of the SGS-term in the $4/5$ law, $G^{1,0}_{L,L}(r,\Delta)$. As noticed, we observe a deviation from the MF prediction and a good agreement with the {\it purely  dissipative} scaling $G^{1,0}_{L,L}(r,\Delta) \sim r ^{\zeta_2-1}$. This can be understood considering that using the  Smagorinsky closure
(\ref{eq:smagorinsky}), one breaks the phase correlation between the three velocity increments entering in the SGS modelling, due to the fact that two terms appear inside the square-root and have a definite sign. As a result, concerning multi-scale correlation, the SGS Smagorinsky stress will  behave as $\tau^{\Delta,SMAG} \sim const\,  \bar s_{ij}$. If this is the case, one predicts the SGS tensor to act as a linear dissipative operator: 
\begin{equation}
\label{eq:smag}
G^{n,0}_{L,L}(r,\Delta) \sim \partial_r D^{n+1,0}(r) \sim r ^{\zeta_{n+1}-1}
\end{equation}
explaining the scaling shown in Fig.(\ref{fig:smag-fourfifth-fullDNS}) for $G^{1,0}_{L.L}(r,\Delta)/r \sim r^{\zeta_2-2}$. 

\noindent
In summary, the Smagorinsky SGS-model performs well at the level of the 
second-order balance equation in the sense that: \\
(1) Both $D^{3,0}(r)$ and $G^{1,0}_{L,L}(r,\Delta)$ obtained from the LES 
show the same scaling behaviour as those obtained from filtered DNS. \\
(2) The effect of the SGS-stress on the two-point energy balance is subleading. \\
(3) The measured scaling of $G^{1,0}_{L,L}(r,\Delta)$ obtained from the Smagorinsky  LES is robust under increasing scale separation between $r$ and $\Delta$.\\
(4) At the same resolution, the LES simulation have a larger extension of the DNS, even if  compared with the hyperviscous Navier-Stokes case.   \\

\subsection{Higher-order balances $(3\leqslant n \leqslant 6)$}
\begin{figure}
\includegraphics[width=0.9\textwidth]{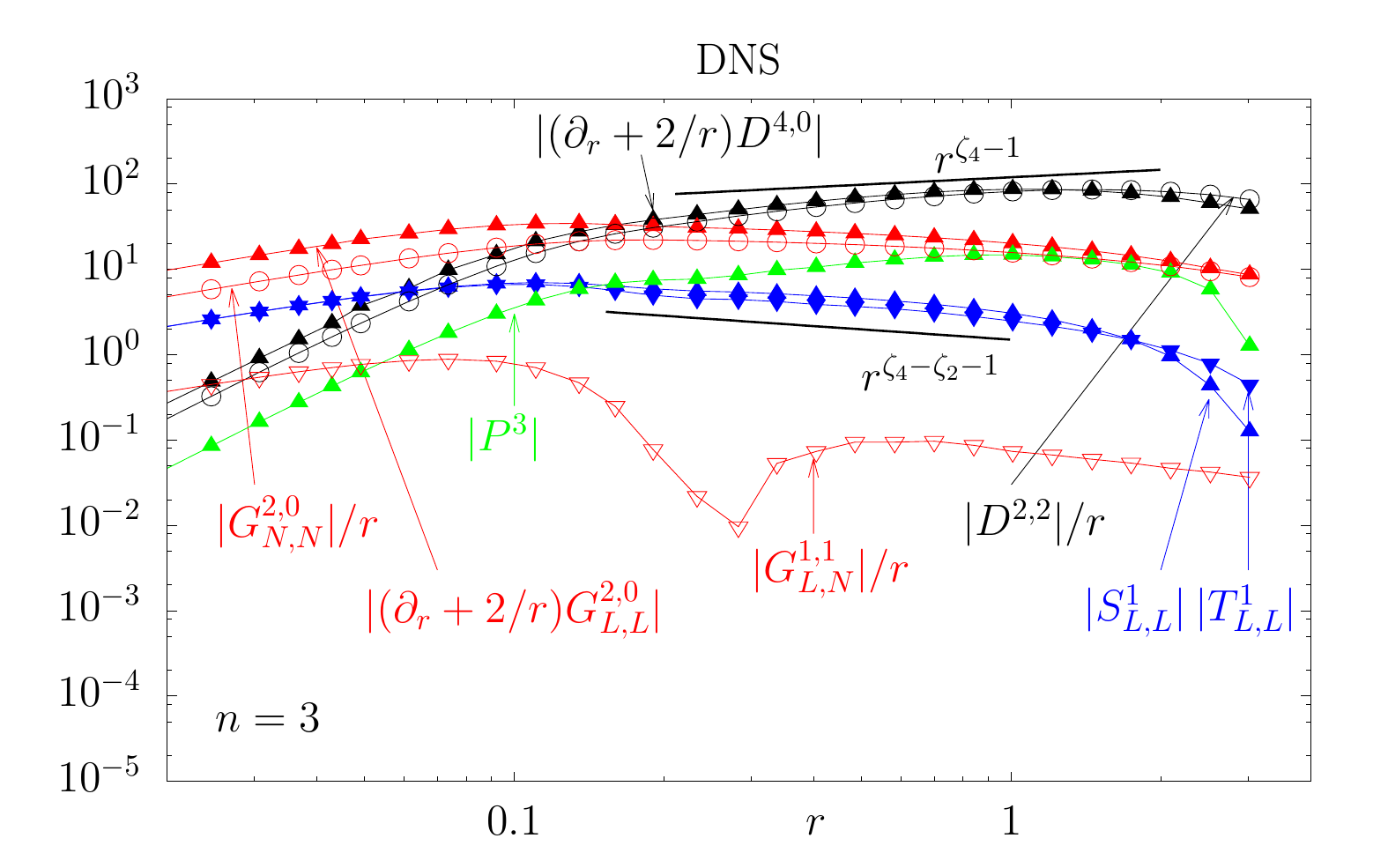} \\ 
\includegraphics[width=0.9\textwidth]{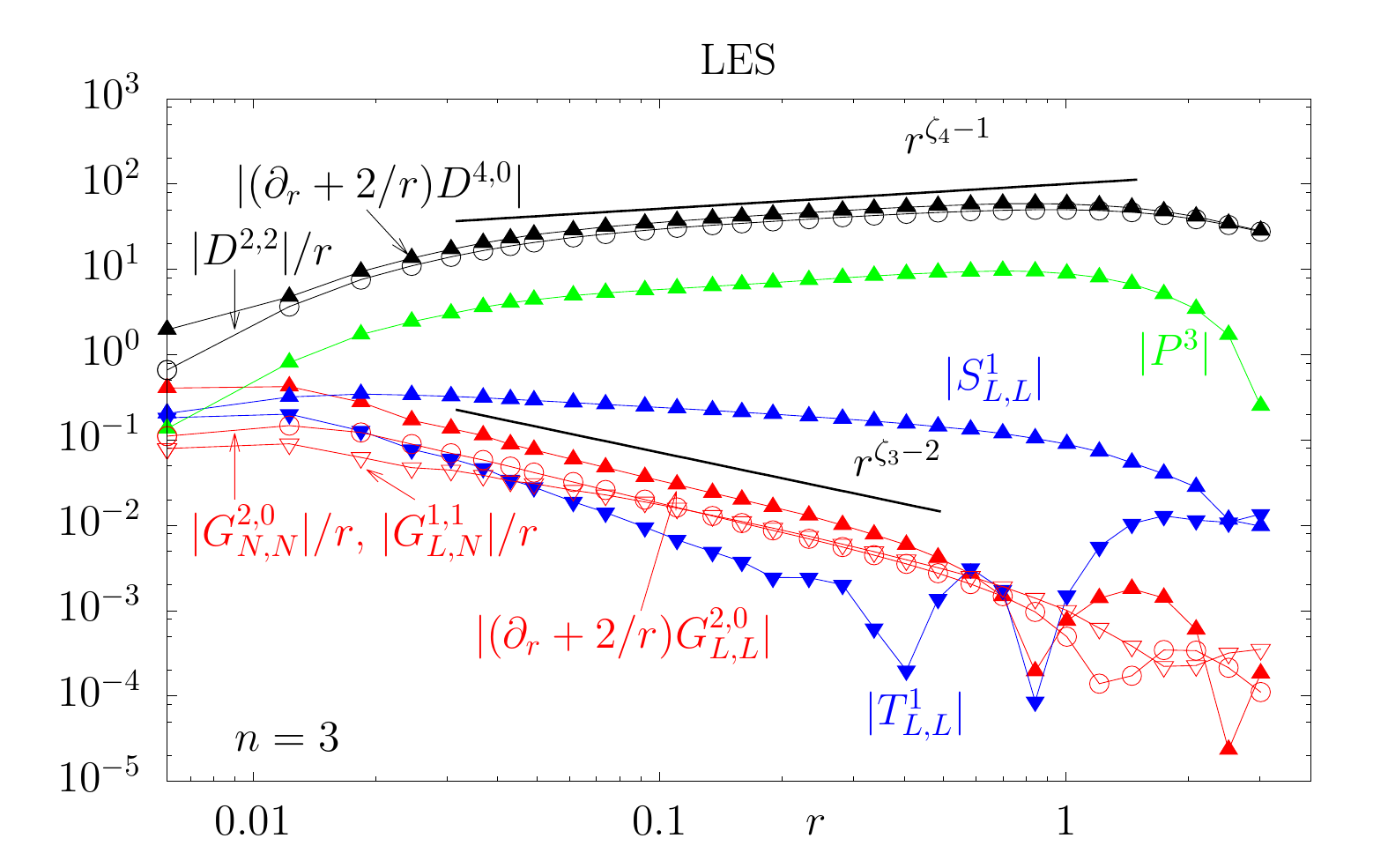}
\caption{
Absolute value of terms contributing to the third-order balance equations, explicitly written in the right-hand side of eq.~\eqref{eq:n3}, for filtered DNS at $\Delta = \pi/40$ from
data-set H1  (top panel) and for Smagorinsky LES from 
data-set LES3  (bottom panel). 
}
\label{fig:higher-order-balance-3}
\end{figure}

\begin{figure}
\centering
\includegraphics[width=0.48\textwidth]{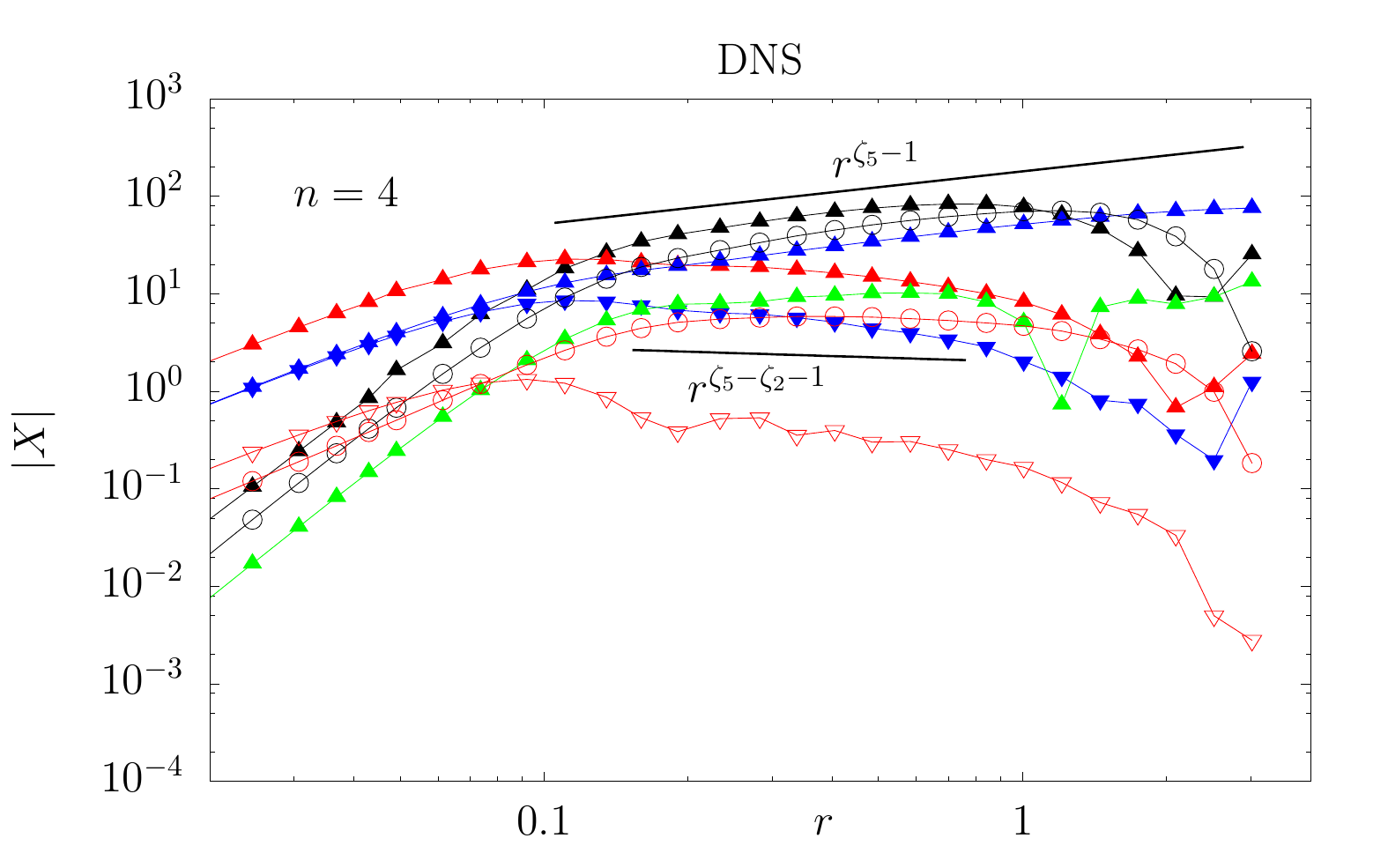}\hspace{1em}
\includegraphics[width=0.48\textwidth]{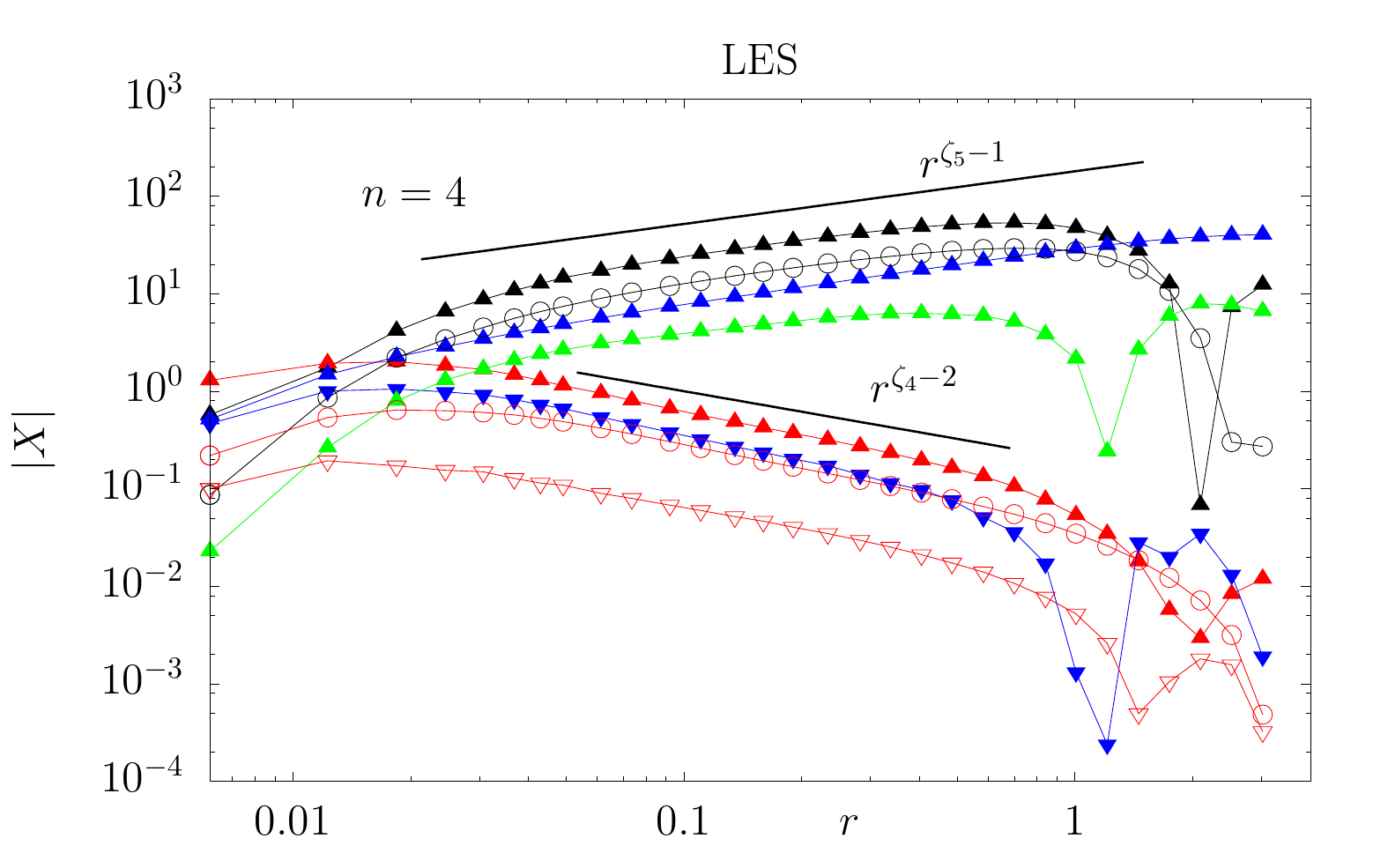}
\caption{
Absolute value of all terms, $|X|$, contributing to the fourth-order balance equations, explicitly written in the right-hand side of eq.~\eqref{eq:n4}, for filtered DNS at $\Delta = \pi/40$ from
data-set H1  (left panel) and for Smagorinsky LES from 
data-set LES3  (right panel).
Black: $| \left(\p_r+\frac{2}{r}\right) D^{5,0}(r)|$ (solid triangles), $|\left(\frac{8}{r}\right)\ D^{3,2}(r)|$ (circles).
Red/grey: $|\left(\p_r+\frac{2}{r}\right) G^{3,0}_{L,L}(r,\Delta)|$ (solid triangles), 
     $|\left(\frac{16}{r}\right) G^{3,0}_{N,N}(r,\Delta)|$ (circles).
     $|\left(\frac{16}{r}\right) G^{2,1}_{L,N}(r,\Delta)|$ (hollow triangles).
Blue/dark grey: $|S^{2}_{L,L}(r,\Delta)|$ (solid up triangles), $|T^{2}_{L,L}(r,\Delta)|$ (solid down triangles). 
Green/light grey: $|P^{4}(r,\Delta)|$. 
}
\label{fig:higher-order-balance-4}
\end{figure}

\begin{figure}
\centering
\includegraphics[width=0.48\textwidth]{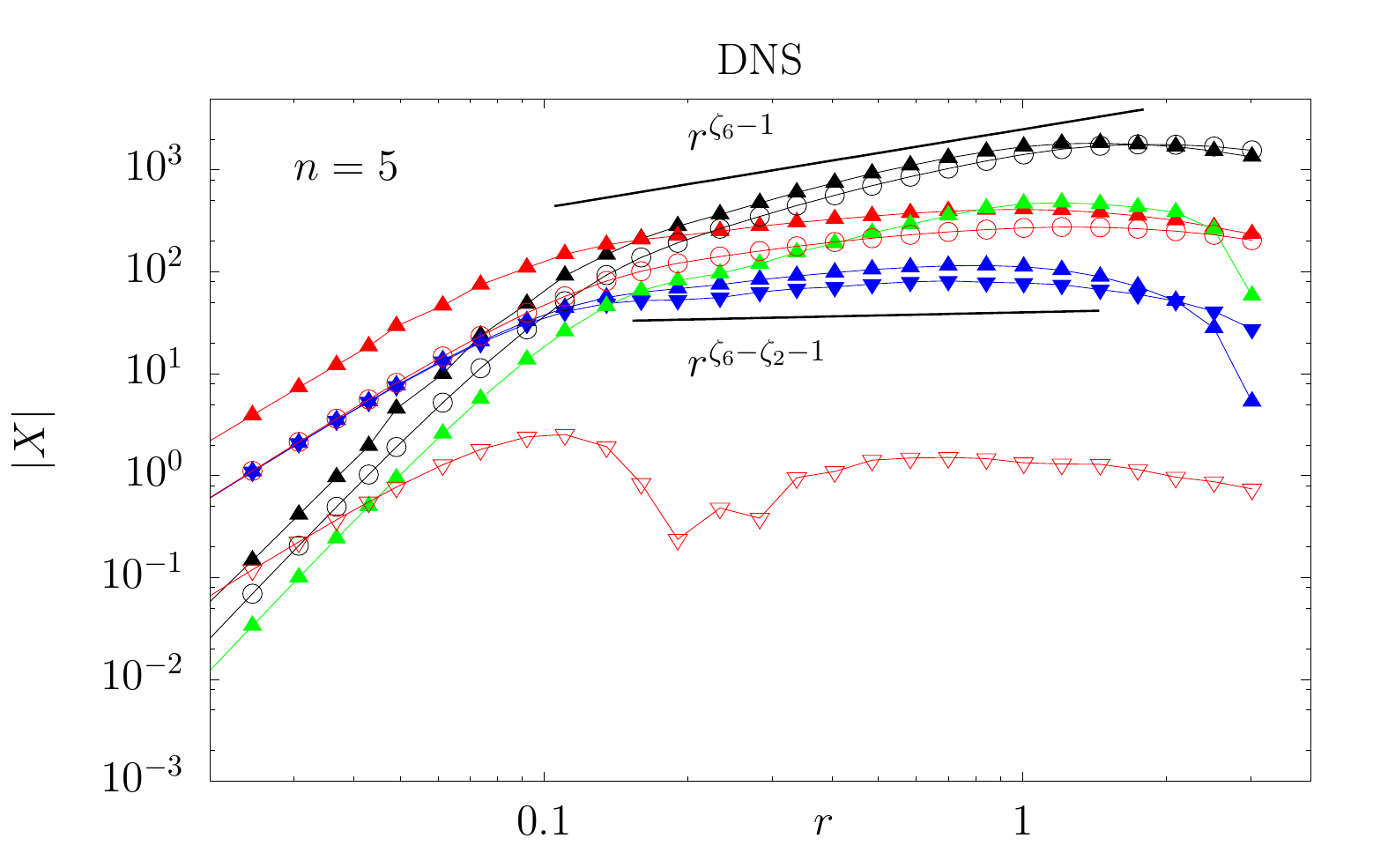}\hspace{1em}
\includegraphics[width=0.48\textwidth]{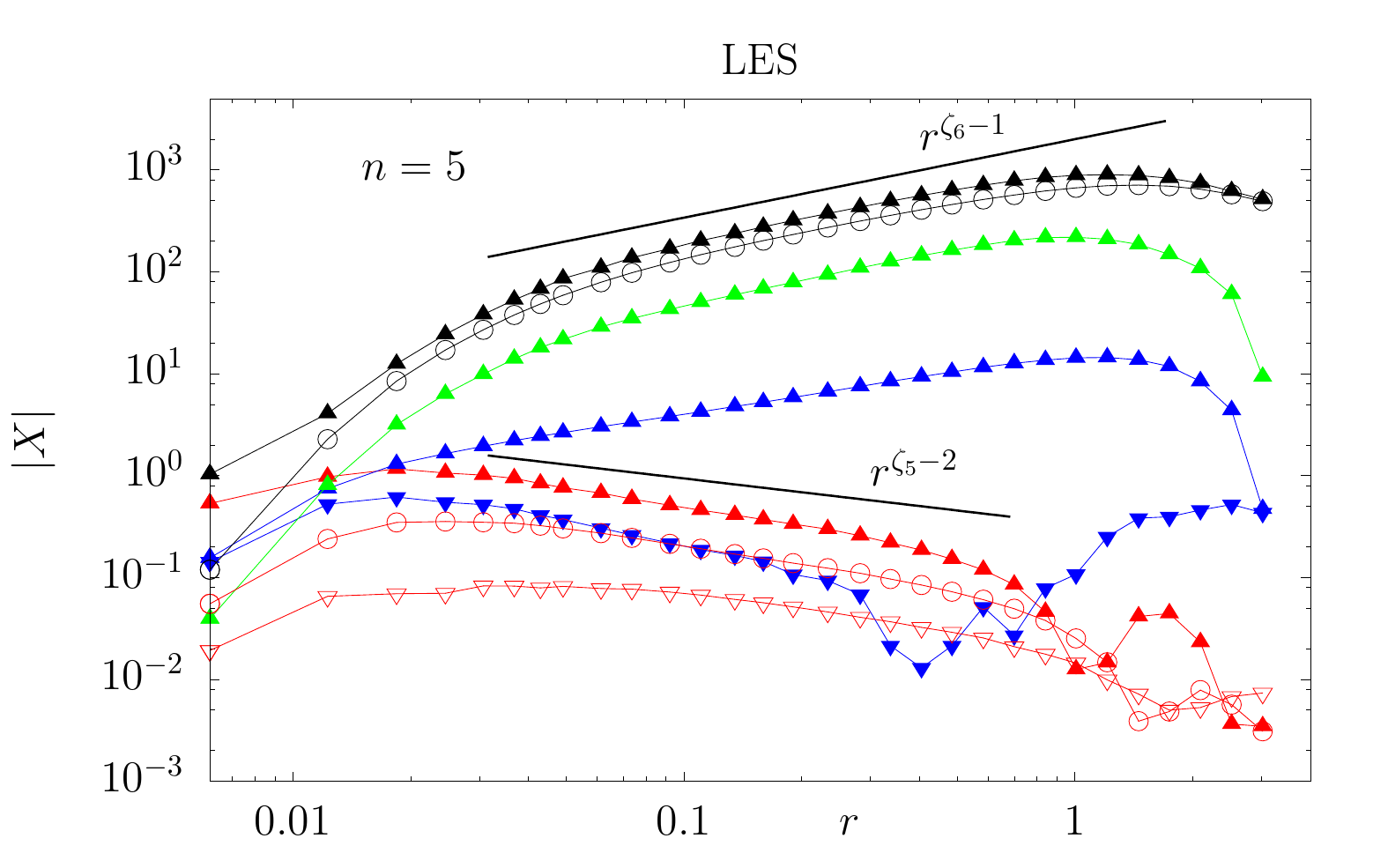}
\caption{
Absolute value of all terms, $|X|$, contributing to the fifth-order balance equations, coming from the right hand side of the hierarchy eq.~\eqref{eq:hierarchy} for $n=5$. (Left panel) data for filtered DNS at $\Delta = \pi/40$ from data-set H1 and (right panel) data for Smagorinsky LES from 
data-set LES3.
Black: $| \left(\p_r+\frac{2}{r}\right) D^{6,0}(r)|$ (solid triangles), $|\left(\frac{10}{r}\right)\ D^{4,2}(r)|$ (circles).
Red/grey: $|\left(\p_r+\frac{2}{r}\right) G^{4,0}_{L,L}(r,\Delta)|$ (solid triangles), 
     $|\left(\frac{20}{r}\right) G^{4,0}_{N,N}(r,\Delta)|$ (circles).
     $|\left(\frac{20}{r}\right) G^{3,1}_{L,N}(r,\Delta)|$ (hollow triangles).
Blue/dark grey: $|S^{3}_{L,L}(r,\Delta)|$ (solid up triangles), $|T^{3}_{L,L}(r,\Delta)|$ (solid down triangles). 
Green/light grey: $|P^{5}(r,\Delta)|$. 
}
\label{fig:higher-order-balance-5}
\end{figure}

\begin{figure}
\centering
\includegraphics[width=0.48\textwidth]{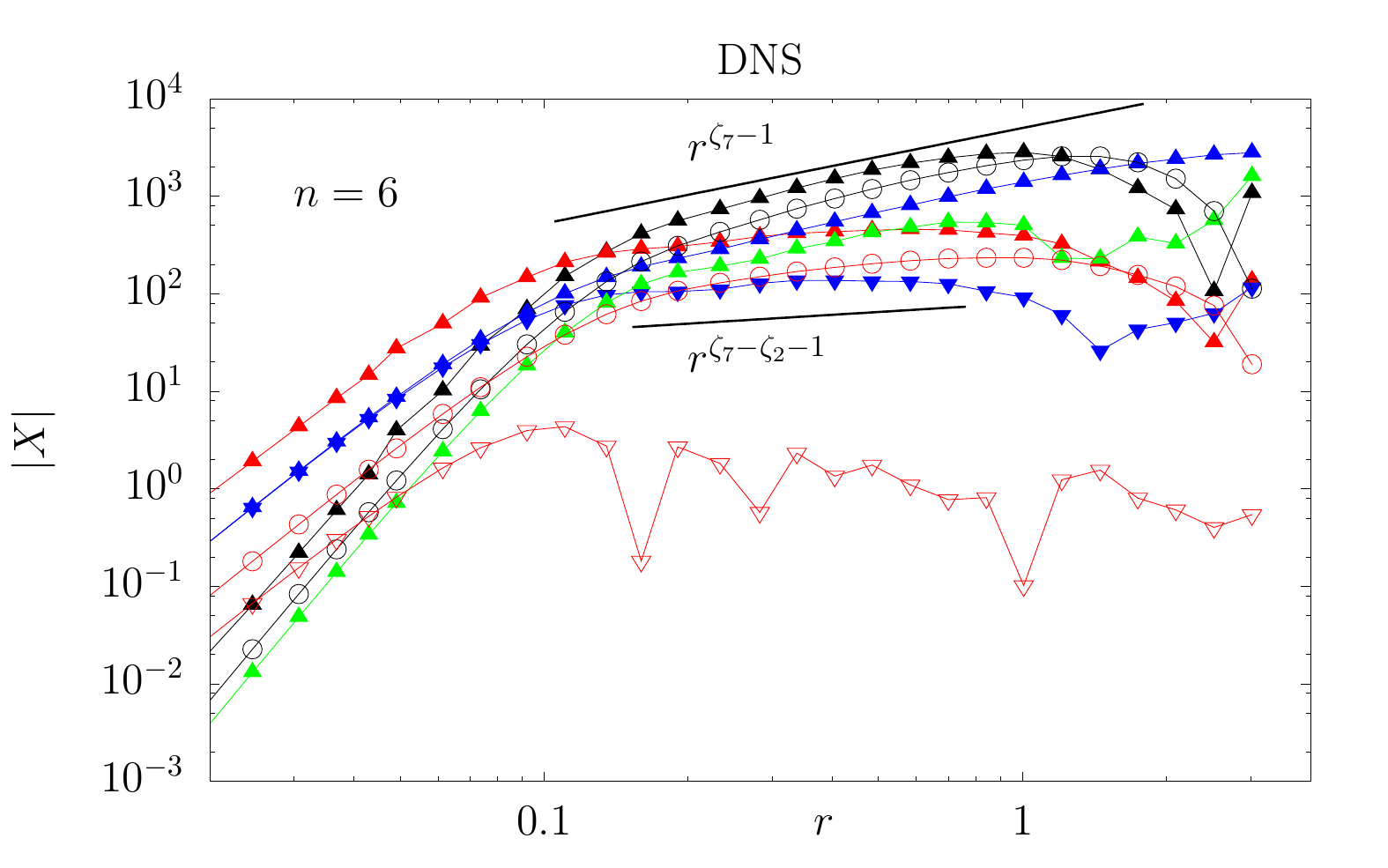}\hspace{1em}
\includegraphics[width=0.48\textwidth]{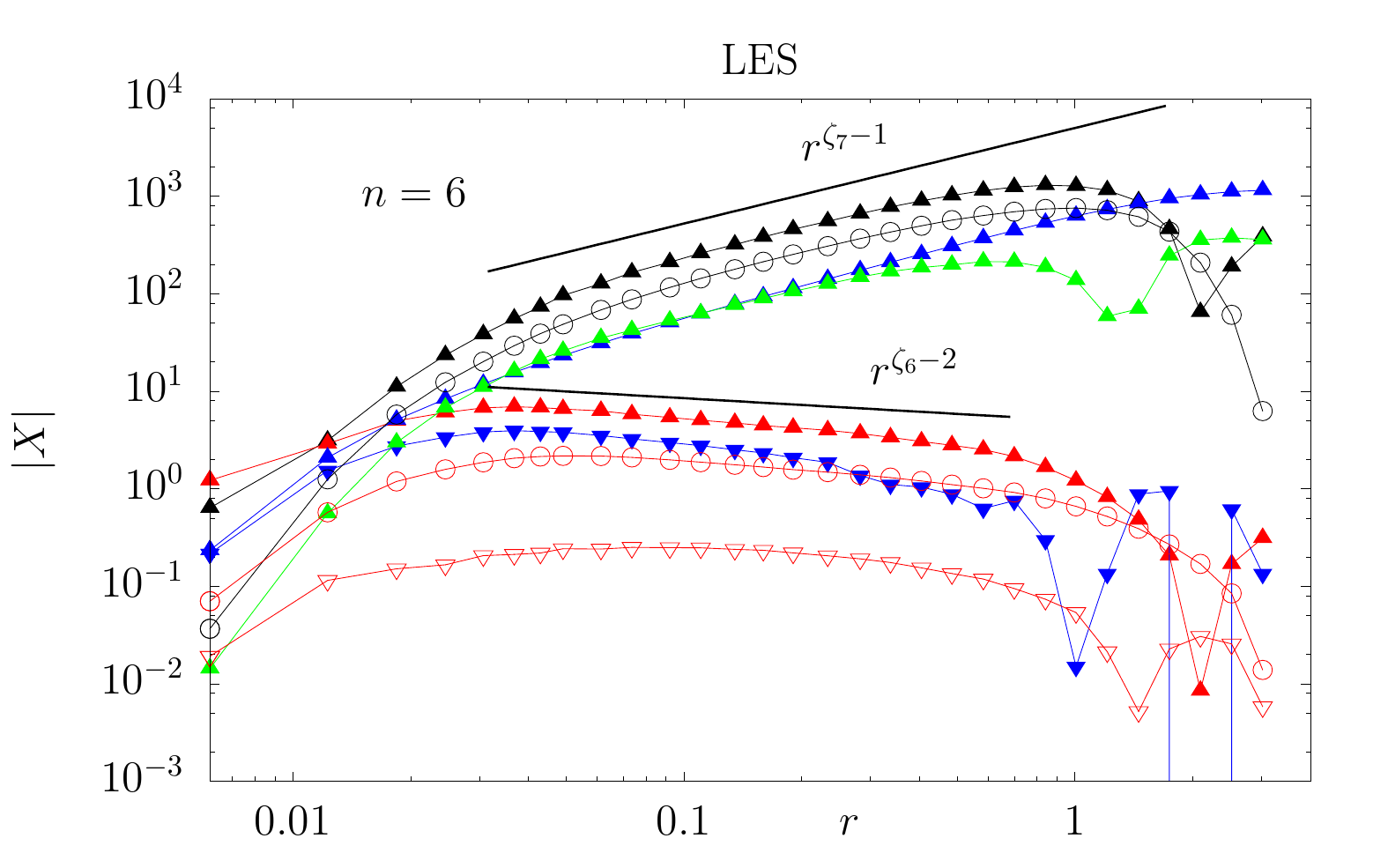}
\caption{Absolute value of all terms, $|X|$, contributing to the sixth-order balance equations, coming from the right hand side of the hierarchy eq.~\eqref{eq:hierarchy} for $n=6$. (Left panel) data for filtered DNS at $\Delta = \pi/40$ from data-set H1 and (right panel) data for Smagorinsky LES from data-set LES3.
Black: $| \left(\p_r+\frac{2}{r}\right) D^{7,0}(r)|$ (solid triangles), $|\left(\frac{12}{r}\right)\ D^{5,2}(r)|$ (circles).
Red/grey: $|\left(\p_r+\frac{2}{r}\right) G^{5,0}_{L,L}(r,\Delta)|$ (solid triangles), 
     $|\left(\frac{24}{r}\right) G^{5,0}_{N,N}(r,\Delta)|$ (circles).
     $|\left(\frac{24}{r}\right) G^{4,1}_{L,N}(r,\Delta)|$ (hollow triangles).
Blue/dark grey: $|S^{4}_{L,L}(r,\Delta)|$ (solid up triangles), $|T^{4}_{L,L}(r,\Delta)|$ (solid down triangles). 
Green/light grey: $|P^{6}(r,\Delta)|$. 
}
\label{fig:higher-order-balance-6}
\end{figure}

\noindent
Having examined the properties of correlations between velocity field  increments and the SGS-stress at the lowest nontrivial 
order in the LES structure function hierarchy, we now examine the higher-order balances.  Here, we need  to study 
the correlations which involve the resolved velocity field 
gradients, $T^n_{i,j}(r,\Delta)$ and $S^n_{i,j}(r,\Delta)$ also. 
The latter describe the correlations between velocity field increments  and part of the SGS energy transfer. As will become clear in the following, even- and odd-order balances require separate descriptions. 
In Figure \ref{fig:higher-order-balance-3} top panel we present all  terms in Eq.~\eqref{eq:n3}
for $n=3$, obtained from the filtered DNS data-set 
H1 for $k_c=40$, while Figure \ref{fig:higher-order-balance-3} bottom panel presents the same terms  for the {\em a-posteriori} data-set LES3. 
The higher-order analysis, $n=4$, $n=5$ and $n=6$, are reported, respectively, in  Figures \ref{fig:higher-order-balance-4}-\ref{fig:higher-order-balance-6} where the results from filtered DNS are presented in the left panels and compared to the one from the LES3 data-set shown in the right panels. The forcing term is not shown in order to improve the readability of the individual figures.\\
Let us comment the general trends. \\
{\bf (i)} For all orders,  {\em a-priori} (DNS) and {\em a-posteriori} (LES) data are in  pretty good agreement, especially concerning the leading terms. This is seen by noticing that for all orders the inertial range behaviour is dominated by the  structure functions, $D^{n,m}(r)$ (black data). Moreover, the scaling is in agreement with the MF prediction (\ref{eq:Dmf}) and LES data do scale better than DNS data. \\
{\bf (ii)} Correlation function involving Pressure (green/light grey data) do scale similarly to $D^{n,m}(r)$, suggesting a key role of them in the global balance. \\
{\bf (iii)} For even orders $(n=4,6)$ also the correlation, $S^{n}_{i,j}(r,\Delta)$ involving the SGS-energy transfer (blue/dark grey colour in Fig. \ref{fig:higher-order-balance-4} and Fig.\ref{fig:higher-order-balance-6}  ) play a leading role, in agreement with what was found for the equivalent terms involving the correlation with the energy dissipation in Navier-Stokes case in \cite{Boschung17}. 
\\
{\bf (iv) } The ensemble of correlation involving the SGS tensor $G^{n,m}(r,\Delta)$, (red/grey data) are always sub-leading and  DNS data do show a different scaling from LES data.\\
{\bf (v)} Correlation given by the terms, $T^n_{i,j}(r,\Delta)$,  are never leading with respect to $S^n_{i,j}(r,\Delta)$  (both in blue/dark grey colour in all Figures) as argued after Eq. (\ref{eq:pi-scaling-estimate-casc}). \\
Let us now comment more on the previous results. 
We first focus on the analysis of the data from the filtered DNS. As can be seen from a qualitative comparison the odd- and even-order balances show important differences. 
For the odd orders (Figures \ref{fig:higher-order-balance-3} and \ref{fig:higher-order-balance-5}),
the pressure correlations  must balance the inertial Structure Functions contribution, $D^{n,m}(r)$,  as all the other terms scale 
in a sub-leading way. For the even orders (Figures \ref{fig:higher-order-balance-4} and \ref{fig:higher-order-balance-6}),
the inertial terms are balanced also by the terms $S^n_{i,j}(r,\Delta)$, which describe the correlations between  longitudinal velocity-field increments and  the components of the SGS-energy  transfer. These differences could have been expected from  numerical results concerning the hierarchy of structure functions in the original NSE obtained in \cite{Boschung17}. Indeed, similar to the present case of filtered DNS, it was found in \cite{Boschung17} that the inertial 
contributions are balanced by the pressure for the odd-order balances.  For the even-order balances,  the inertial contributions are balanced also by the contributions from  the viscous terms. The latter is similar to our results for filtered DNS, as the  correlations between the resolved-scale velocity increments and the SGS energy transfer play a similar role 
to the viscous energy dissipation in the full Navier-Stokes evolution -- with the important difference that energy dissipation is point-wise positive definite in the NSE. 
Differences between even- and odd-order balances, in the filtered DNS data, are also visible concerning 
  the functions $S^n_{i,j}(r,\Delta)$ and $T^n_{i,j}(r,\Delta)$. The latter is always decaying by going to larger and large scale separations, $r/\Delta \gg1 $, the former  matches the MF prediction (\ref{eq:MFleading}) only for even order (see right panels of Figs. (\ref{fig:higher-order-balance-4}) and  (\ref{fig:higher-order-balance-6}), while odd orders are much more depleted and very close to $T^n_{i,j}(r,\Delta)$ (top panel of Fig.~\ref{fig:higher-order-balance-3} and left panel of Fig.~\ref{fig:higher-order-balance-5}). The above behaviour can be understood by noticing that odd-order $S^n_{i,j}(r,\Delta)$ correlations involves unsigned velocity increments and SGS energy transfer, introducing non-trivial cancellations that brings the quantity away from its leading MF prediction.\\ Concerning  LES data, we found that $S^n_{i,j}(r,\Delta)$ is in good agreement with DNS for even orders (left columns of Figs. (\ref{fig:higher-order-balance-4}) and  (\ref{fig:higher-order-balance-6}), while it is more intense than the {\em a-priori} case for odd orders (bottom panel of Fig.~\ref{fig:higher-order-balance-3} and right panel of Fig.~\ref{fig:higher-order-balance-5}). This is probably due to the fact that in the Smagorinsky LES the SGS energy transfer is positive definite,  and it is not able to reproduce the cancellations present in the real DNS, leading to a contribution larger than what would be in reality. 
\noindent
To be more quantitative, we show in all figures the straight-line corresponding to scaling MF predictions (\ref{eq:MFleading})  for the dominant contribution, which is in very good agreement for all cases.  
The values for the scaling exponents $\zeta_n$ of the $n^{\rm th}$-order 
longitudinal correlation functions used in this comparison are taken from Ref.~\cite{Gotoh02,benzi2010inertial}, i.e., $\zeta_2 = 0.70 \pm 0.01$, $\zeta_4 = 1.29 \pm 0.03$, $\zeta_5 = 1.54 \pm 0.03$, $\zeta_6 = 1.77 \pm 0.04$ 
and $\zeta_7 = 1.98 \pm 0.06$. \\
As noticed, the whole set of multi-scale correlation function involving the SGS stress (red data)  given by the 
class $G(r,\Delta)$ are always sub-leading with respect to $D^{n,m}(r)$ and to the pressure and they are in good agreement with the MF prediction (\ref{eq:MFsubleading}) for the {\em a-priori} DNS data and with (\ref{eq:smag}) for the {\em a-posteriori} LES case (as shown by the corresponding straight lines in all plots). 
\noindent 
Before concluding, let us summarise the main findings. \\
\noindent {\bf 1.} A simple LES approach based on a Smagorinsky model is able to reproduce most of the multi-scale physical properties of real turbulence at high Reynolds numbers, including the MF scaling in the inertial range of the structure functions, $D^{n,m}(r)$,  and of the correlation among velocity increments and the SGS energy transfer, $S^n(r,\Delta)$ (for even $n$).\\
\noindent {\bf 2.}  Nevertheless,  some notable differences arise. In particular, the multi-scale correlations (\ref{eq:G}) involving the SGS stress tensor have a smaller amplitude in LES than for real DNS. By comparing the left and right columns of Figs. (\ref{fig:higher-order-balance-4}-\ref{fig:higher-order-balance-6}) one clearly sees that for LES data, there exists a sharp  difference among those correlations that have the leading scaling behaviour and those that follow off for large $r/\Delta$ separation. This fact is  a {\it positive} outcome, indicating that the LES closure has a minor influences on the inertial range scaling properties than standard viscosity or hyperviscous effects. A different trend is also measured for the behaviour of correlation, $S^3_{ij}(r,\Delta)$, which appears in the balance for $n=5$ in  Fig. (\ref{fig:higher-order-balance-5}) where the LES data (right panel) do have a better scaling with a larger exponents than the DNS data (left panel). 
\\
\section{Conclusions}
\label{sec:conclusions}
This paper provides analytical and numerical results concerning the multiscale correlations between the resolved-scale velocity field  increments and SGS quantities for {\em a-priori} and {\em a-posteriori} data. We derived the exact hierarchy of higher-order equations for all structure  functions obtained from filtered NSE.  All correlations were measured  using a database consisting of filtered DNS  on up to $2048^3$ and Smagorinsky LES on up to $1024^3$ collocation points. \\
Under the assumption of a connection between resolved-scale and SGS statistics given by  a multiplicative MF cascade process,  we provided scaling estimates for all two-point  functions involving correlations among the resolved velocity increments and the SGS stress or the SGS energy transfer.  
\\
Concerning the comparison between filtered DNS and Smagorinsky LES, we find that the results obtained from the Smagorinsky model agree well with those from filtered DNS concerning 
all leading terms, i.e.  those involving  structure functions, pressure correlations and correlations with the SGS-energy transfer. On the contrary, all terms involving correlations with SGS tensor, $G_{i,j}^{n,m}(r,\Delta)$ have a smaller amplitude and a faster decrease as a function of the scale separation $\Delta/r$  for   LES data.   \\
Overall, the LES approach works well, leading to a larger extension of the scaling range with respect to the DNS at comparable numerical resolution. 
\blue{
Since the Smagorinsky model performs well concerning the scaling of the 
structure functions, in principle we do not expect significant improvements from
more sophisticated models.  
However, a better LES model may be more efficient in                       
reproducing inertial-range scaling than the Smagorinsky model, in the sense
that coarser grids may be possible for a suitable model.}
A more detailed quantitative assessment of the effects of LES modelling on the inertial range scaling properties, also comparing different subgrid closures will be presented elsewhere. \\
\noindent
{Moreover,} the numerical study of the {\em a-priori} filtered DNS data revealed some differences in the scaling behaviour of correlation functions belonging to even- and odd-order balance equations, 
similar to results concerning the viscous contributions obtained from the  full Navier-Stokes evolution \cite{Boschung17}. For even-order balances the leading terms are the ones given by velocity structure functions, $D^{n,m}(r)$, the pressure, $P^n(r,\Delta)$ and the correlation with the SGS energy transfer, $S^n_{i,j}(r,\Delta)$. All of them follow the MF prediction (\ref{eq:MFleading}). For odd orders, the cross correlation involving the SGS energy balance are subleading.  Terms involving the cross correlation with the SGS tensor, $G_{i,j}^{n,m}(r,\Delta)$ are always sub-leading as well, as predicted by the MF; see Eq.~(\ref{eq:MFsubleading}). \\ 
In the Appendices 
we reproduce all technical aspects concerning the derivation of the exact hierarchy for LES 
{(see Appendix \ref{app:derivation})} and for a modification of the closure equations  where we explicitly took into account also the re-projection 
of the nonlinear term involving the two  filtered fields  \cite{Pope00,carati2001,winckelmans2001,Buzzicotti17a} 
which is somehow unavoidable in numerical applications; 
{see Appendix \ref{app:ples-hierarchy}}. The latter induced complications in the derivation of the structure function hierarchy, which are tackled through a distinction between the actual 
SGS stress and the Leonard stress \cite{Leonard75}. The contributions from the Leonard stress only appear in the higher-order balance equation, with the four-fifth law ($n=2$ in the hierarchy) remaining unaffected (see Appendix \ref{app:ples-hierarchy} and Ref.~\cite{Buzzicotti17a}).   
{The properties of the correlation tensors, which are used in the 
derivation of the LES-hierarchy of longitudinal structure 
functions are summarised in Appendices \ref{app:gtensors} and \ref{app:stensors} for those of type $G_{i,j}^{n,m}(r,\Delta)$, $S_{i,j}^{n}(r,\Delta)$ and $T_{i,j}^{n}(r,\Delta)$, where we also
comment on the general structure of these terms. The latter is 
also used in Appendix \ref{app:ptensors} in order to provide an explicit form of the pressure correlation, which distinguishes 
correlations involving velocity-field gradients from those only involving correlations between the pressure and the velocity-field
increments. Finally, Appendix \ref{app:fourfifth} contains a 
re-derivation of the $4/5$-th law for LES from the tensorial 
approach, and a subsequent comparison to the corresponding result
in Ref.~\cite{Meneveau94}.} 
{
It is important to stress  that studies similar to the one presented here can be performed also in the presence of anisotropy, for wall-bounded \cite{casciola2003scale} and high-Reynolds boundary layer flows \cite{meneveau2013generalized}. In these cases, the injection of energy due to the coupling with the mean shear leads to an increasing of  intermittency and to a more complicated  scale-by-scale energy balance \cite{perot1995shear, marati2004energy}. Small-scale vorticity production is the key mechanisms that  needs to be  captured by LES acting prominently at the cut-off scales \cite{leveque2007shear,cui2007new}. Feedback of intense-but-rare small-scale fluctuations  on the resolved large-scale and on the mean profiles is even more important than in  homogeneous and isotropic case. An extension of our present study on LES of wall bounded flows, would help to better quantify the accuracy of different sub-grid models for such systems also for  higher-orders statistics.
}

\section*{Acknowledgements}
We acknowledge useful discussions with H. Aluie, R. Benzi, J. Brasseur and C. Meneveau. The research leading to these results has received funding from the European
Union's Seventh Framework Programme (FP7/2007-2013) under grant agreement No. 339032.

\appendix

\section{Derivation of the LES structure function hierarchies}
\label{app:derivation}
This appendix contains the derivation of Eq.~\eqref{eq:hierarchy}
from Eq.~\eqref{eq:les_increment_evol}. 
Since all tensors in Eq.~\eqref{eq:les_increment_evol} 
which do not contain explicit correlations to the SGS stress
are structurally identical to those figuring in the evolution equations derived from the
full NSE, the only term that needs to be considered is the tensor
\be
\label{eq:H-tensor}
H_{i_{1} \hdots i_{n}} =
\frac{1}{|S_{n-1}|}\sum_{\sigma \in S_n}
   \langle \delta_r \ov_{i_{\sigma(1)}} \hdots \delta_r \ov_{i_{\sigma(n-1)}} \delta_r (\p_k\tau^\Delta_{k i_{\sigma(n)}}) \ .
\ee
This expression has been obtained through point splitting, that is
one considers two points $\bx'$ and $\bx$ such that $\bx' = \bx + \br$.
In order to separate single- and multiscale contributions to the
tensor $H$, we carry out a change of variables using
\begin{align}
\bX & =  \frac{1}{2} (\bx + \bx') \ \mbox{ and } \ \br = \bx'- \bx \ , \\
\bx' & = \bX + \frac{1}{2} \br \ \mbox{ and } \ \bx = \bX - \frac{1}{2} \br \ ,
\end{align}
which leads to
\begin{align}
\label{eq:partial_X}
\p_{X_i} &= \p_{x_i'} + \p_{x_i} \ , \\
\label{eq:partial_r}
\p_{r_i} &= \frac{1}{2}(\p_{x_i'} - \p_{x_i})\ , \\
\label{eq:partial_xpr}
\p_{x_i'}& = \frac{1}{2}\p_{X_i} + \p_{r_i} \ , \\
\label{eq:partial_x}
\p_{x_i} &= \frac{1}{2}\p_{X_i} - \p_{r_i} \ .
\end{align}
In the new coordinates the increment $\delta_r (\p_k\tau^\Delta_{ki})$ can be written as
\be
\label{eq:tau_increment}
\delta_r (\p_k\tau^\Delta_{ki}) = \p_{x_k'}\tau^{\Delta \prime}_{ki} - \p_{x_k}\tau^\Delta_{ki}
= (\p_{x_k'}-\p_{x_k})(\tau^{\Delta \prime}_{ki} +\tau^\Delta_{ki}) =  2\p_{r_k}(\tau^{\Delta \prime}_{ki} + \tau^\Delta_{ki}) \ .
\ee

Substitution of this equation into the expression for $H_{i_{1} \hdots i_{n}}$ in eq.~\eqref{eq:H-tensor} yields
\begin{align} 
H_{i_{1} \hdots i_{n}}& =  
\frac{1}{|S_{n-1}|}\sum_{\sigma \in S_n} 
   \langle \delta_r \ov_{i_{\sigma(1)}} \hdots \delta_r \ov_{i_{\sigma(n-1)}} \delta_r (\p_k\tau^\Delta_{k i_{\sigma(n)}}) \rangle  \nonumber \\
& = \frac{2}{|S_{n-1}|}\sum_{\sigma \in S_n} 
   \langle \delta_r \ov_{i_{\sigma(1)}} \hdots \delta_r \ov_{i_{\sigma(n-1)}} 
           \p_{r_k} (\tau^{\Delta \prime}_{ki_{\sigma(n)}} + \tau^\Delta_{ki_{\sigma(n)}}) \rangle   \nonumber \\
& = \frac{2}{|S_{n-1}|}\sum_{\sigma \in S_n} \p_{r_k} 
    \langle \delta_r \ov_{i_{\sigma(1)}} \hdots \delta_r \ov_{i_{\sigma(n-1)}} 
           (\tau^{\Delta \prime}_{ki_{\sigma(n)}} + \tau^\Delta_{ki_{\sigma(n)}}) \rangle \nonumber \\ 
& \hspace{-4em} - \frac{2}{|S_{n-1}||S_{n-2}|}\sum_{s \in S_{n-1}}\sum_{\sigma \in S_n} 
    \langle \delta_r \ov_{i_{s(\sigma(1))}} \hdots \delta_r \ov_{i_{s(\sigma(n-2))}}
    \p_{r_k}(\delta_r \ov_{i_{s(\sigma(n-1))}})(\tau^{\Delta \prime}_{ki_{\sigma(n)}} + \tau^\Delta_{ki_{\sigma(n)}}) \rangle \ .
\end{align}
The summands in the last term on the RHS of this equation can also be written as
\begin{align}
& \langle \delta_r \ov_{i_{s(\sigma(1))}} \hdots \delta_r \ov_{i_{s(\sigma(n-2))}}
      \p_{r_k}(\delta_r \ov_{i_{s(\sigma(n-1))}})(\tau^{\Delta \prime}_{ki_{\sigma(n)}} + \tau^\Delta_{ki_{\sigma(n)}}) \rangle \nonumber \\
& \ \ = \frac{1}{2}\langle \delta_r \ov_{i_{s(\sigma(1))}} \hdots \delta_r \ov_{i_{s(\sigma(n-2))}}
     (\p_{x_k'}\ov_{i_{s(\sigma(n-1))}}' + \p_{x_k}\ov_{i_{s(\sigma(n-1))}})(\tau^{\Delta \prime}_{ki_{\sigma(n)}} + \tau^\Delta_{ki_{\sigma(n)}}) \rangle \ , 
\end{align}
such that
\begin{align} 
H_{i_{1} \hdots i_{n}}& =  
 \frac{2}{|S_{n-1}|}\sum_{\sigma \in S_n} \p_{r_k} \langle \delta_r \ov_{i_{\sigma(1)}} \hdots \delta_r \ov_{i_{\sigma(n-1)}} 
                                                    (\tau^{\Delta \prime}_{ki_{\sigma(n)}} + \tau^\Delta_{ki_{\sigma(n)}}) \rangle \nonumber \\ 
& \ \ - \frac{1}{|S_{n-1}||S_{n-2}|}\sum_{s \in S_{n-1}}\sum_{\sigma \in S_n} \nonumber \\ 
    & \qquad     \langle \delta_r \ov_{i_{s(\sigma(1))}} \hdots \delta_r \ov_{i_{s(\sigma(n-2))}}
     (\p_{x_k'}\ov_{i_{s(\sigma(n-1))}}' + \p_{x_k}\ov_{i_{s(\sigma(n-1))}})(\tau^{\Delta \prime}_{ki_{\sigma(n)}} + \tau^\Delta_{ki_{\sigma(n)}}) \rangle 
\nonumber \\
& = \frac{2}{|S_{n-1}|}\sum_{\sigma \in S_n} \p_{r_k} \langle \delta_r \ov_{i_{\sigma(1)}} \hdots \delta_r \ov_{i_{\sigma(n-1)}} 
                                                    (\tau^{\Delta \prime}_{ki_{\sigma(n)}} + \tau^\Delta_{ki_{\sigma(n)}}) \rangle \nonumber \\ 
& \ \ - \frac{1}{|S_{n-1}||S_{n-2}|}\sum_{s \in S_{n-1}}\sum_{\sigma \in S_n} \nonumber \\
     & \qquad    \langle \delta_r \ov_{i_{s(\sigma(1))}} \hdots \delta_r \ov_{i_{s(\sigma(n-2))}}
     (\p_{x_k'}\ov_{i_{s(\sigma(n-1))}}'\tau^{\Delta \prime}_{ki_{\sigma(n)}} + \p_{x_k}\ov_{i_{s(\sigma(n-1))}}\tau^\Delta_{ki_{\sigma(n)}}) \rangle 
\nonumber \\
& \ \ - \frac{1}{|S_{n-1}||S_{n-2}|}\sum_{s \in S_{n-1}}\sum_{\sigma \in S_n} \nonumber \\
     & \qquad  \langle \delta_r \ov_{i_{s(\sigma(1))}} \hdots \delta_r \ov_{i_{s(\sigma(n-2))}}
     (\p_{x_k'}(\ov_{i_{s(\sigma(n-1))}}'\tau^\Delta_{ki_{\sigma(n)}}) + \p_{x_k}(\ov_{i_{s(\sigma(n-1))}}\tau^{\Delta \prime}_{ki_{\sigma(n)}}) \rangle \ , 
\end{align}
and we have separated three contributions; the first term on the RHS describes the correlation between the velocity
field increments and the SGS tensor, while the second term describes the correlations between the
velocity field increments with the velocity field gradients and the SGS tensor evaluated at the same point and
the third term describes the correlations between the
velocity field increments with the field gradients and the SGS tensor evaluated at different points.
For $n=2$ the second term becomes the subgrid energy flux.
We define three tensors to keep track of the different correlations
\begin{align}
G_{i_{1} \hdots i_{n}k}(r,\Delta) 
& = \langle \delta_r \ov_{i_1} \hdots \delta_r \ov_{i_{n-1}}(\tau^{\Delta \prime}_{i_n k} + \tau^\Delta_{i_nk}) \rangle \ , \\ 
S_{i_{1} \hdots i_{n}}(r,\Delta) & 
\equiv \langle \delta_r \ov_{i_1} \hdots \delta_r \ov_{i_{n-2}}
     (\p_{x_k'}\ov_{i_{n-1}}' \tau^{\Delta \prime}_{i_nk} + \p_{x_k}\ov_{i_{n-1}}\tau^{\Delta}_{i_nk}) \rangle \ , \\
T_{i_{1} \hdots i_{n}}(r,\Delta) & 
\equiv \langle \delta_r \ov_{i_1} \hdots \delta_r \ov_{i_{n-2}}
     (\p_{x_k'}\ov_{i_{n-1}}' \tau^{\Delta}_{i_nk} + \p_{x_k'}\ov_{i_{n-1}}'\tau^{\Delta}_{i_nk}) \rangle \ ,
\end{align}
and introduce their symmetrised versions
\begin{align}
G_{\{i_{1} \hdots i_{n}\}k}(r,\Delta) &\equiv \frac{1}{|S_{n-1}|}\sum_{\sigma \in S_n} G_{i_{\sigma(1)} \hdots i_{\sigma(n)}k}(r,\Delta) \ , \\
S_{\{i_{1} \hdots i_{n}\}}(r,\Delta) & \equiv \frac{1}{|S_{n-1}||S_{n-2}|}\sum_{s \in S_{n-1}}\sum_{\sigma \in S_n} 
         S_{i_{s(\sigma(1))} \hdots i_{s(\sigma(n-1))i_{\sigma(n)}}}(r,\Delta) \ , \\
T_{\{i_{1} \hdots i_{n}\}}(r,\Delta) & \equiv \frac{1}{|S_{n-1}||S_{n-2}|}\sum_{s \in S_{n-1}}\sum_{\sigma \in S_n} 
         T_{i_{s(\sigma(1))} \hdots i_{s(\sigma(n-1))i_{\sigma(n)}}}(r,\Delta) \ .
\end{align}
We can therefore express the tensor $H_{i_{1} \hdots i_{n}}$ as follows
\be
H_{i_{1} \hdots i_{n}} = 2 \p_{r_k} G_{\{i_{1} \hdots i_{n}\}k} - S_{\{i_{1} \hdots i_{n}\}} - T_{\{i_{1} \hdots i_{n}\}} \ .
\ee
From their definitions, it is clear that $S_{\{i_{1} \hdots i_{n}\}}$ and
$T_{\{i_{1} \hdots i_{n}\}}$ are isotropic tensors which are symmetric under the
exchange of any pair of indices, the same applies to $\p_{r_k} G_{\{i_{1} \hdots i_{n}\}k}$. Therefore
$H_{i_{1} \hdots i_{n}}$ is an isotropic tensor which is symmetric under the exchange of any two indices, as it
must be.
Hence eq.~\eqref{eq:les_increment_evol} can be written more concisely as
\begin{align}
\p_t \langle \delta_r \ov_{i_1} \hdots \delta_r \ov_{i_n} \rangle  = & 
- \p_{r_k} \langle \delta_r \ov_{i_1} \hdots \delta_r \ov_{i_n} \delta_r \ov_k \rangle
- P_{\{i_{1} \hdots i_{n}\}} \nonumber \\
 & - 2 \p_{r_k} G_{\{i_{1} \hdots i_{n}\}k}(r,\Delta) + S_{\{i_{1} \hdots i_{n}\}}(r,\Delta) + T_{\{i_{1} \hdots i_{n}\}}(r,\Delta) + F_{\{i_{1} \hdots i_{n}\}}\ , 
\end{align}
where $P_{\{i_{1} \hdots i_{n}\}}$ denotes the correlation tensor between the velocity and
pressure gradient increments and $F_{\{i_{1} \hdots i_{n}\}}$ the correlation with the force increments.
Note that the pressure tensor is structurally similar to $H_{i_{1} \hdots i_{n}}$;
therefore, a similar splitting should be possible (see also Ref.~\cite{Hill01_arx}) and may be interesting
in order to extend the results of Ref.~\cite{Hill01} by inclusion of the
pressure-velocity correlation functions in explicit form.

The divergence of arbitrary $n^{th}$-order isotropic tensors which are symmetric under the
exchange of two indices was calculated in general in Ref.~\cite{Hill01} with details given in
Ref.~\cite{Hill01_arx}. These results can now be applied here, leading to the following
hierarchy of equations for the $n$-th order longitudinal structure function $D^{n,0}$:
\begin{align}
\label{eqapp:hierarchy}
\partial_t D^{n,0}(r) =& - \left( \p_r  D^{n+1,0}(r) + \frac{2}{r} D^{n+1,0}(r) - \frac{2n}{r} D^{n-1,2}(r) \right) \nonumber \\
& - 2n\left(\p_r +\frac{2}{r}  \right) G^{n-1,0}_{L,L}(r,\Delta)  + \frac{4n}{r}(G^{n-2,1}_{L,N}(r,\Delta)+ G^{n-1,0}_{N,N}(r,\Delta))\nonumber \\
& +  \frac{2n!}{(n-2)!}(S^{n-2}_{L,L}(r,\Delta)  + T^{n-2}_{L,L}(r,\Delta)) - 2nP^n(r,\Delta)  + 2nF^n(r)\ , 
\end{align}
where the usual choice $\br = (r,0,0)$ was used and the divergence of the $G$-tensors has been evaluated,
leading to the presence of the functions $G^{n-1,0}_{L,L}(r,\Delta)$, $G^{n-2,1}_{L,N}(r,\Delta)$ and 
$G^{n-1,0}_{N,N}(r,\Delta)$. The function 
$F^n$ denotes the contribution from the forcing
\be
F^n = \langle \underbrace{\delta_r \ov_L \hdots \delta_r \ov_L}_\text{n-1 times}
                                             \delta_r f_L \rangle \ .
\ee
In order to derive this final hierarchy of equations,
the tensors $S$ and $T$ and the divergence of the tensors $G$ must be evaluated.
Unlike the tensors involving only velocity increments, the tensors $G$ are in general not symmetric with respect to the exchange of arbitrary pairs of indices, which precludes the direct application of results
from Ref.~\cite{Hill01}. Details of the evaluation of $\p_{r_k} G_{\{i_{1} \hdots i_{n}\}k}$ can be found 
in Appendix \ref{app:gtensors}, and the evaluation of the tensors of type $S$ and $T$ is carried out in appendix 
\ref{app:stensors}. The pressure tensors are considered in appendix \ref{app:ptensors}.

\subsection{Recovery of the four-fifth law for LES for $n=2$}
\label{app:fourfifth}
We now treat the longitudinal components of $S_{\{ij\}}$, $T_{\{ij\}}$ and $(\p_{r_k}G_{\{ij\}k})$
on the RHS of the tensor equation for the longitudinal case
more in detail in order to relate eq.~\eqref{eq:n2} to the corresponding result in Ref.~\cite{Meneveau94}.
We begin by evaluating $(\p_{r_k}G_{\{ij\}k})$. From the definition of the third-order tensor
\be
G_{\{ij\}k} = \langle \delta_r \ov_i (\tau^{\Delta \prime}_{jk} + \tau^\Delta_{jk})\rangle
            + \langle \delta_r \ov_j (\tau^{\Delta \prime}_{ki} + \tau^\Delta_{ki})\rangle \ ,
\ee
we obtain
\begin{align}
G_{\{ij\}k} & =
2 \left[\langle \ov_i' \tau^\Delta_{jk})\rangle + \langle \ov_j'\tau^\Delta_{ki}\rangle \right] \ , 
\end{align}
since $\langle \ov_i \tau^{\Delta \prime}_{jk}\rangle = -\langle \ov_i' \tau^\Delta_{jk}\rangle$ (see Appendix \ref{app:gtensors}).
The evaluation of the divergence of $G_{\{ij\}k}$ can be simplified through the incompressibility constraint, which 
results in $\p_{r_k} \langle \ov_k'\tau^\Delta_{ij}\rangle =0$. Therefore one obtains 
\begin{align}
\p_{r_k} G_{\{ijk\}}(r,\Delta)
& = 2 \p_{r_k} \langle \ov_i'\tau^\Delta_{jk} + \ov_j'\tau^\Delta_{ik} + \ov_k'\tau^\Delta_{ij} \rangle 
\nonumber \\
& = 2 \p_{r_k} \langle \ov_i'\tau^\Delta_{jk} + \ov_j'\tau^\Delta_{ik} \rangle 
= \p_k G_{\{ij\}k} \ , 
\end{align}
where the tensor
$G_{\{ijk\}} = G_{\{ij\}k} + \langle \delta_r \ov_k (\tau^{\Delta \prime}_{ij} + \tau^\Delta_{ij})\rangle$ is an isotropic
tensor which is symmetric under exchange of any pair of indices. 
Alongside incompressibility, these geometric constraints result in $G_{\{ijk\}}$ to be of the following form  
\cite{Robertson40,Monin75}
\be
G_{\{ijk\}} = (G^{1,0}_{L,L} -r \p_r G^{1,0}_{L,L}) \frac{3r_i r_j r_k}{2r^3} 
+ (G^{1,0}_{L,L} +r \p_r G^{1,0}_{L,L})\left[\frac{r_i}{2r} \delta_{jk} + \frac{r_j}{2r} \delta_{ki} + \frac{r_k}{2r} \delta_{ij} \right] \ ,
\ee
and its divergence can be calculated using the general results on the divergence of
an isotropic tensor which is symmetric under exchange of any pair of indices (see Ref.~\cite{Hill01})
\begin{align}
\p_{r_k} G_{\{ijk\}}
& = \left[ \left(\p_r + \frac{2}{r}\right) \left(\frac{3(G^{1,0}_{L,L}-r\p_r G^{1,0}_{L,L})}{2}\right) + \left(2\p_r - \frac{2}{r}\right) 
     \left(\frac{G^{1,0}_{L,L}+r\p_r G^{1,0}_{L,L}}{2}\right)  \right] \frac{r_ir_j}{r^2}  
\nonumber \\
 & \ \ +  \left[ \left(\p_r + \frac{4}{r}\right) \left(\frac{G^{1,0}_{L,L}+r\p_r G^{1,0}_{L,L}}{2}\right)\right]\delta_{ij} \ .
\end{align}
The evaluation of the tensor $S_{\{ij\}}$ is straightforward
\begin{align}
\label{eq:stensor_2ndorder}
S_{\{ij\}} & =
\langle (\p_{x'_k} \ov_i') \tau^{\Delta \prime}_{jk} + (\p_{x_k} \ov_i) \tau^\Delta_{jk} +
            (\p_{x'_k} \ov_j') \tau^{\Delta \prime}_{ki} + (\p_{x_k} \ov_j) \tau^\Delta_{ki}\rangle \nonumber \\
& = 2 \langle (\p_{x_k} \ov_i) \tau^\Delta_{jk} + (\p_{x_k} \ov_j) \tau^\Delta_{ik} \rangle  \ .
\end{align}
Owing to the incompressibility constraint, the tensor $T_{\{ij\}}$ can in fact be expressed in terms of 
the divergence of $G$ 
\begin{align}
\label{eq:ttensor_2ndorder}
T_{\{ij\}} & =
\langle (\p_{x'_k} \ov_i' + \p_{x_k} \ov_i) (\tau^{\Delta \prime}_{jk} + \tau^\Delta_{jk})\rangle
            + \langle (\p_{x'_k} \ov_j' + \p_{x_k} \ov_j) (\tau^{\Delta \prime}_{ki} + \tau^\Delta_{ki})\rangle \nonumber \\
& = \langle \p_{x_k} (\ov_i\tau^{\Delta \prime}_{jk}) + \p_{x'_k} (\ov_i'\tau^\Delta_{jk})\rangle
 + \langle \p_{x_k} (\ov_j\tau^{\Delta \prime}_{ik}) + \p_{x'_k} (\ov_j'\tau^\Delta_{ik})\rangle 
\nonumber \\
& =2 \p_{r_k} \langle \ov_i'\tau^\Delta_{jk} + \ov_j'\tau^\Delta_{ik}  \rangle 
= \p_{r_k} G_{\{ijk\}}(r,\Delta)  \ ,
\end{align}
where the last equality follows from incompressibility: $\p_{r_k} \langle \ov_k'\tau^\Delta_{ij}\rangle =0$. 
The longitudinal components of three tensors then become
\begin{align}
\left( \p_{r_k} G_{\{ijk\}}(r,\Delta) \right)_{LL} &= 2 \frac{\p_r}{r^4} (r^4G^{1,0}_{L,L}(r,\Delta)) = \frac{\p_r}{r^4} (r^4G_{LLL}(r,\Delta))\ , \\ 
S^2_{L,L} & = \langle \tau^\Delta_{kL}\partial_{x_k}\ov_L \rangle \ , \\
T^2_{L,L}(r,\Delta) & = \frac{1}{4}\left( \p_{r_k} G_{\{ijk\}}(r,\Delta) \right)_{LL} = \frac{\p_r}{4r^4} (r^4G^{1,0}_{L,L}(r,\Delta)) \ , \\ 
\label{eq:stensor_2ndorder_long}
\end{align}
and we obtain
\be
\p_t D^{2,0}  = - \frac{\p_r}{3r^4}(r^4D^{3,0}) - 2 \frac{\p_r}{r^4} (r^4G^{1,0}_{L,L}) +  4 S^2 +  \frac{\p_r}{r^4} (r^4G^{1,0}_{L,L})  + 2F^2 \ , 
\ee
from which for $\p_t D^{2,0}=0$ and $F^2=0$ we recover eq.~(47) in Ref.~\cite{Meneveau94} stated here in the 
notation used in Ref.~\cite{Meneveau94}
\be
\label{eqapp:2ndorder-Meneveau}
D_{LLL} = -6 G_{L,LL}(r,\Delta) +  \frac{12}{5} \langle \tau^\Delta_{kL}\partial_{x_k}\ov_L \rangle r = -6 G_{L,LL}(r,\Delta) + 6\langle \tau^\Delta_{LL}s_{LL} \rangle r \ .
\ee
where 
\be
G_{LLL}(r,\Delta) = \langle \ov_L'\tau^\Delta_{LL}\rangle
= \frac{1}{2} \langle \delta_r \ov_L (\tau^{\Delta \prime}_{LL}+\tau^\Delta_{LL})\rangle
= \frac{1}{2}G^{1,0}_{L,L}(r,\Delta) \ ,
\ee
and $D_{LLL} \equiv D^{3,0}$.
Concerning the last equality in Eq.~\eqref{eqapp:2ndorder-Meneveau}, 
note that the term $6\langle \tau^\Delta_{LL}s_{LL} \rangle$ must equal the SGS energy flux in stationary state, which implies
$6\langle \tau^\Delta_{LL}s_{LL} \rangle = -\frac{4}{5}\langle \Pi(\Delta) \rangle = \frac{4}{5}\langle \tau^\Delta_{kj}s_{kj} \rangle$, which
implies $\langle \tau^\Delta_{LL}s_{LL} \rangle = \frac{2}{15}\langle \tau^\Delta_{kj}s_{kj} \rangle$ 
(see also eq.~ (71) \cite{Meneveau94}). 
These relations also imply $\langle \tau^\Delta_{kL}\partial_{x_k}\ov_L \rangle = \langle \tau^\Delta_{kj}s_{kj} \rangle/3$.
The contribution from the forcing gives
\be
2F^2 = 2 \langle \delta_r \ov_L \delta_r f_L \rangle = \frac{4}{3}\eps_{\rm IN}-4\langle \ov_L'f_L\rangle \ ,
\ee
and for $r << L_f$ we can approximate $\langle \ov_L'f_L\rangle \simeq \eps_{\rm IN}/3$
(by isotropy $\langle \ov_L f_L\rangle = \eps_{\rm IN}/3$).
Hence in stationary state, where $\langle \Pi \rangle = \eps_{\rm IN}$, we again
recover the four-fifth law with $\langle \Pi \rangle$ replaced by the numerically equal $\eps_{\rm IN}$.

\section{Projected LES}
\label{app:ples-hierarchy}
In LES applications, the coarse computational grid cannot resolve 
small-scale dynamics generated by the coupling of resolved-scale velocity-field
components. In formal terms, the existence of the LES grid therefore requires 
the evolution of the resolved-scale velocity field to be confined to the
same finite-dimensional vector space; hence, 
it is necessary to project the momentum equation again, resulting in   
\be
\label{eq:momentum-ples}
\p_t \ov_i + \p_j(\overline{\ov_i\ov_j} + \overline{P}\delta_{ij} + \tau^{\Delta,P}_{ij}) = f_i \ .
\ee
The equation derived from this momentum balance at points $\bx$ and $\bx'$ for
the $n^{th}$-order tensor of velocity-field increments in 
homogeneous isotropic turbulence then reads 
\begin{align}
\label{eq:ples_increment_evol}
\p_t \langle \delta_r \ov_{i_1} \hdots \delta_r \ov_{i_n} \rangle  = &    
-\frac{1}{|S_{n-1}|}\sum_{\sigma \in S_n} \nonumber \\
& \left \langle \delta_r \ov_{i_{\sigma(1)}} \hdots \delta_r \ov_{i_{\sigma(n-1)}} 
\left[ \overline{ \left(\frac{\ov_k' + \ov_k}{2} \right)
 \p_{X_k} \delta_r \ov_{i_{\sigma(n)}}} 
+ \overline{\delta_r \ov_k \p_{r_k}\delta_r \ov_{i_{\sigma(n)}}} 
\right] \right \rangle 
\nonumber \\
& - \frac{1}{|S_{n-1}|}\sum_{\sigma \in S_n} 
 \langle \delta_r \ov_{i_{\sigma(1)}} \hdots \delta_r \ov_{i_{\sigma(n-1)}} \delta_r (\p_k\overline{P}\delta_{k i_{\sigma(n)}}) \rangle
\nonumber \\
& - \frac{1}{|S_{n-1}|}\sum_{\sigma \in S_n} \langle \delta_r \ov_{i_{\sigma(1)}} \hdots \delta_r \ov_{i_{\sigma(n-1)}} \delta_r (\p_k\tau^{\Delta,P}_{k i_{\sigma(n)}}) \rangle  \nonumber \\
& + \frac{1}{|S_{n-1}|}\sum_{\sigma \in S_n} \langle \delta_r \ov_{i_{\sigma(1)}} \hdots \delta_r \ov_{i_{\sigma(n-1)}} \delta_r f_{i_{\sigma(n)}} \rangle  \ ,
\end{align}
where 
\be
\tau^{\Delta,P}_{ij} \equiv \overline{v_i v_j}-\overline{\ov_i\ov_j} \ .
\ee 
A further advantage of the projected LES formulation is that 
$\tau^{\Delta,P}_{ij}$ only consists of SGS quantities while the unprojected SGS stress
$\tau^\Delta_{ij}$ includes a residual coupling amongst resolved scales. 
A detailed discussion of the difference between the two formulations is 
given in Refs.~\cite{winckelmans2001,carati2001,Buzzicotti17a}. 
For the derivation of a hierarchy of equations for the 
structure functions from Eq.~\eqref{eq:ples_increment_evol} 
we immediately run into several difficulties:

\begin{enumerate} 
\item The correlation tensor of the velocity field increments which arises from the nonlinear term in 
eq.~\eqref{eq:momentum-ples} is no longer symmetric under the exchange of any two indices.
\item Due to the asymmetry caused by the additional projector acting on the nonlinear term, 
      the derivatives with respect to $X_k$ cannot be removed by homogeneity, because they 
      cannot be brought out from inside the average.   
\item If we wish to bring the derivatives with respect to $r_k$ out from inside the average
      additional terms appear from the product rule of differentiation again due the asymmetry 
      introduced by the projector acting on the nonlinear term.  
\item We cannot relate the higher-order structure functions ($n \geqslant 3$, see below) 
      to each other without the introduction of a correction term.
\end{enumerate}

However, by introducing a correction term 
\be
\label{eq:corrector}
\tau^{\Delta,LEO}_{ij} \equiv \overline{\ov_i\ov_j}-\ov_i\ov_j \ ,
\ee
which is known as the {\em Leonard stress} \cite{Leonard75}, it is possible to 
rewrite the momentum balance \eqref{eq:momentum-ples} as 
\be
\label{eq:momentum-ples-corr}
\p_t \ov_i + \p_j(\ov_i\ov_j + \overline{P}\delta_{ij} + \tau^{\Delta,P}_{ij} + \tau^{\Delta,LEO}_{ij}) = f_i \ . 
\ee
Using the correction terms originating from the Leonard stress, 
we obtain for the evolution of the $n^{th}$-order correlation tensor
of velocity-field increments
\begin{align}
\label{eq:ples_increment_evol_corr}
\p_t \langle \delta_r \ov_{i_1} \hdots \delta_r \ov_{i_n} \rangle  = &    
-\p_{r_k} \langle \delta_r \ov_{i_1} \hdots \delta_r \ov_{i_n} \delta_r \ov_k \rangle  
\nonumber \\
& - \frac{1}{|S_{n-1}|}\sum_{\sigma \in S_n} \langle \delta_r \ov_{i_{\sigma(1)}} \hdots \delta_r \ov_{i_{\sigma(n-1)}} \delta_r (\p_k\overline{P}\delta_{k i_{\sigma(n)}}) \rangle
\nonumber \\
& - \frac{1}{|S_{n-1}|}\sum_{\sigma \in S_n} 
\langle \delta_r \ov_{i_{\sigma(1)}} \hdots \delta_r \ov_{i_{\sigma(n-1)}} \delta_r (\p_k\tau^{\Delta,P}_{k i_{\sigma(n)}}) \rangle  
\nonumber \\
& - \frac{1}{|S_{n-1}|}\sum_{\sigma \in S_n} 
\langle \delta_r \ov_{i_{\sigma(1)}} \hdots \delta_r \ov_{i_{\sigma(n-1)}} \delta_r (\p_k\tau^{\Delta,LEO}_{k i_{\sigma(n)}}) \rangle  \nonumber \\
& + \frac{1}{|S_{n-1}|}\sum_{\sigma \in S_n} \langle \delta_r \ov_{i_{\sigma(1)}} \hdots \delta_r \ov_{i_{\sigma(n-1)}} \delta_r f_{i_{\sigma(n)}} \rangle  \ ,
\end{align}
where the tensors involving $\tau^{\Delta,P}_{ij}$ and $\tau^{\Delta,LEO}_{ij}$ have the same symmetries. Hence, we can use the results 
from the derivation of the equation hierarchy following 
Eq.~\eqref{eq:les_increment_evol} to deduce the corresponding hierarchy for the projected LES (P-LES) 
following Eq.~\eqref{eq:ples_increment_evol_corr}
\begin{align}
\label{eq:ples-allorders}
\partial_t D^{n,0} &= - \left(\p_r + \frac{2}{r}\right) (D^{n+1,0} + 2n (G^{n-1,0,\rm P}_{L,L}+G^{n-1,0, \rm LEO}_{L,L})) \nonumber \\ 
& \qquad + \frac{2n}{r} (D^{n-1,2} +2(G^{n-2,1,\rm P}_{L,N}+ G^{n-1,0,\rm P}_{N,N}+ G^{n-2,1,\rm LEO}_{L,N}+ G^{n-1,0, \rm LEO}_{N,N}))
\nonumber \\
& \qquad +  \frac{2n!}{(n-2)!}\left(S^{n-2,\rm P}_{L,L} + S^{n-2, \rm LEO}_{L,L} + T^{n-2, \rm P}_{L,L} + T^{n-2,\rm LEO}_{L,L}\right) 
- 2nP^{n}  + 2nF^{n}\ . 
\end{align}
Note that no correction terms are present for $n=2$
since 
\begin{align}
G^{\rm LEO}_{ijk}&=2\langle \ov_i'\tau^{\Delta,\rm LEO}_{jk}\rangle 
= 2\langle \ov_i'\overline{\ov_j\ov_k}\rangle - 2\langle \ov_i'\ov_j\ov_k\rangle  
= 2\langle \ov_i'\ov_j\ov_k\rangle - 2\langle \ov_i'\ov_j\ov_k\rangle =0 \ , 
\nonumber \\
T^{0,\rm LEO}_{L,L} & = \frac{1}{4}\left(\partial_{r_k} G^{\rm LEO}_{ijk}\right)_{LL} = 0 \ , 
\nonumber \\
S^{0,\rm LEO}_{L,L} &= \langle \tau^{\Delta,\rm LEO}_{kL} \partial_{x_k}\ov_L\rangle = 
\langle (\overline{\ov_k \ov_L}-\ov_k\ov_L)\partial_{x_k}\ov_L \rangle
=\langle \overline{\ov_k \ov_L}\partial_{x_k}\ov_L \rangle
-\langle \ov_k\ov_L\partial_{x_k}\ov_L \rangle =0 \ ,
\end{align}
i.e.~at the level of the third-order correlation function the 
correlation between the resolved field 
and the correction term vanishes. 
This is not the case at higher orders. {Correlation functions
involving the Leonard stress are briefly discussed in Appendix \ref{app:correlations-leo}}.

\subsection{Correlations involving the Leonard stress}
\label{app:correlations-leo}
In order to cover all contributions to the higher-order balance
equations, we briefly describe the correlations involving the 
Leonard stress. As mentioned earlier, these are correction terms
given in terms of resolved-scale quantities and as such do not 
require modelling.  
At all orders $n>2$ in the {\em a-priori} analysis of 
data-set H1 we find that the functions 
$G^{n-1,0, \rm LEO}_{L,L}$, $G^{n-1,0, \rm LEO}_{N,N}$ and $G^{n-2,1, \rm LEO}_{L,L}$
change sign around $\Delta$ and display power-law scaling in the inertial subrange.
Since their scaling exponents are always smaller than those
of the functions $G^{n-1,0, \rm LEO}_{L,L}$
and $G^{n-1,0, \rm LEO}_{N,N}$, 
the correlations involving 
the Leonard stress could in principle become more 
important than those involving 
the actual SGS stress. At all orders considered here, we find that 
the Leonard-stress correlations are small compared 
to the correlations with the SGS stress, however, 
their significance increases in the higher-order equations. 
In contrast, the functions $S^{n,\rm LEO}$, 
which encode correlations
between resolved-scale velocity-field increments and parts of the 
energy transfer amongst the resolved scales, we consistently 
find {\em negative} inertial-range scaling exponents. 
Furthermore, the $S^{n,\rm LEO}$ is always small compared to $S^{n, \rm P}$
at all orders. Hence the contributions of 
$S^{n,\rm LEO}$ to the higher-order energy balances are always 
subleading in the inertial range. 

\section{Properties of correlation tensors}
In this appendix we summarise the properties of the correlation tensors that have been 
used in the derivation of Eq.~\eqref{eq:hierarchy} as outlined in Appendix \ref{app:derivation}.
\subsection{Tensors involving velocity field increments}
\label{app:gtensors}
The tensors
\begin{align}
G_{i_1\hdots i_{n-1},i_nk} = \langle \delta_r \ov_{i_1} \hdots \delta_r \ov_{i_{n-1}} 
(\tau^{\Delta \prime}_{i_n k} + \tau^\Delta_{i_n k}) \rangle \ ,
\end{align}
are isotropic tensors which are symmetric under the exchange of any pair of indices $i_1, \hdots, i_{n-1}$ as well 
as under the exchange of $i_n$ with $k$. That is, it is the tensor product of two tensors 
which are symmetric under the exchange of any pair of indices: a $N=n-1$ subtensor and a $N=2$ subtensor. This 
structure is used to obtain a general formula for the divergence of $G_{i_1\hdots i_{n-1},i_nk}$ based 
on the results of Ref.~\cite{Hill01} for tensors which are symmetric under the exchange of any pair of indices.
Here we provide some more detail on this procedure for the lowest orders and we summarise some useful 
properties of the $G$-tensors.
\subsubsection{Second order correlation tensor}
For $n=2$ we obtain
\begin{align}
\langle \ov_k'\tau^\Delta_{ij}\rangle & = \left(\langle \ov_L'\tau^\Delta_{LL}\rangle -r\p_r \langle \ov_L'\tau^\Delta_{LL}\rangle \right)
\frac{r_i r_j r_k}{2r^3} \nonumber \\
                                  & \quad + \left(2\langle \ov_L'\tau^\Delta_{LL}\rangle +r\p_r \langle \ov_L'\tau^\Delta_{LL}\rangle \right) \left(\frac{r_i}{4r}\delta_{ij} + \frac{r_j}{4r}\delta_{ik}\right) 
                                 - \langle \ov_L'\tau^\Delta_{LL}\rangle \frac{r_k}{2r}\delta_{ij} \ .
\label{eqapp:2ndorderGtensor}  
\end{align}
From eq.~\eqref{eqapp:2ndorderGtensor}, the following property of 
$\langle \ov_k'\tau^\Delta_{ij}\rangle$ can be derived
\be
\langle \ov_k\tau^{\Delta \prime}_{ij}\rangle = \langle \ov_k(\bx)\tau^\Delta_{ij}(\bx + \br)\rangle 
= \langle \ov_k(\bx-\br)\tau^\Delta_{ij}(\bx)\rangle = -\langle \ov_k'\tau^{\Delta}_{ij}\rangle \ ,
\ee
where the second equality follows from homogeneity. This relation further implies that in the limit 
$r \to 0$ one obtains 
\be
\langle \ov_k\tau^\Delta_{ij}\rangle = \lim_{r \to 0} \langle \ov_k\tau^{\Delta \prime}_{ij}\rangle = -\lim_{r \to 0} \langle \ov_k'\tau^\Delta_{ij}\rangle 
= -\langle \ov_k\tau^\Delta_{ij}\rangle \ , 
\ee 
and hence $\langle \ov_k\tau^\Delta_{ij}\rangle = 0$. 
Hence for the correlation function $G^{1,0}_{L,L} = \langle \delta_r \ov_L (\tau^{\Delta \prime}_{LL}+\tau^\Delta_{LL})\rangle$ we obtain
\begin{align}
\langle \delta_r \ov_L \tau^\Delta_{LL}\rangle &
= \langle \delta_r \ov_L (\tau^{\Delta \prime}_{LL}+\tau^\Delta_{LL})\rangle  
-\langle \delta_r \ov_L \tau^{\Delta \prime}_{LL}\rangle
= 2\langle \ov_L'\tau^\Delta_{LL} \rangle- \langle \ov_L'\tau^{\Delta \prime}_{LL}\rangle + \langle \ov_L\tau^{\Delta \prime}_{LL}\rangle 
\nonumber \\
&= \langle \ov_L'\tau^\Delta_{LL} \rangle- \langle \ov_L\tau^\Delta_{LL}\rangle 
= \langle \ov_L'\tau^\Delta_{LL}\rangle = \frac{1}{2}\langle \delta_r \ov_L (\tau^{\Delta \prime}_{LL}+\tau^\Delta_{LL})\rangle  \ . 
\end{align}
The behaviour of the correlation function $G^{1,0}_{L,L}$ in the limit $r \to 0$ can also be 
obtained from 
\be
\frac{1}{2} \partial_r G^{1,0}_{L,L}|_{r=0} = \partial_r \langle \ov_L(\bx + r \hat{\bx})\tau^\Delta_{LL}(\bx)\rangle|_{r=0}
= \langle \partial_r \ov_L'\tau^\Delta_{LL}\rangle|_{r=0} 
= \langle s_{LL}\tau^\Delta_{LL} \rangle = -\frac{2}{15}\langle \Pi(\Delta) \rangle \ , 
\ee
see Ref.~\cite{Meneveau94} or Appendix \ref{app:stensors} for the last equality. Therefore, 
$G^{1,0}_{L,L} = 2\langle s_{LL}\tau^\Delta_{LL} \rangle r$ for small $r$, and $\lim_{r \to 0} G^{1,0}_{L,L}(r) = 0$.

\subsubsection{Third-order correlation tensor}
For $n=3$, we obtain from symmetry considerations for any isotropic tensor invariant
under the pairwise exchange of $i$ with $j$ and $k$ with $l$ \cite{Robertson40,Monin75}
\begin{align}
G_{ij,lk} & = \langle \delta_r \ov_{i} \delta_r \ov_j (\tau^{\Delta \prime}_{l k} + \tau^\Delta_{l k}) \nonumber \\
& = A_1(r,\Delta) r_i r_j r_k r_l + B_1(r,\Delta) r_i r_j \delta_{kl} + B_2(r,\Delta) r_k r_l \delta_{ij} 
\nonumber \\
& + B_3(r,\Delta)(r_i r_k \delta_{jl} + r_i r_l \delta_{jk} + r_k r_j \delta_{il} +  r_j r_l \delta_{ik}) \nonumber \\
&  + C_1(r,\Delta) (\delta_{ik}\delta_{jl} + \delta_{il}\delta_{jk} )  + C_2(r,\Delta)\delta_{ij}\delta_{kl} \ .
\end{align}
For $\br = (r,0,0)$, the only non-zero components are:
\begin{align}
G^{2,0}_{L,L}(r,\Delta) = G^{2,0,0}_{L,L}(r,\Delta) &= A_1 r^4 + (B_1 +B_2 + 4B_3)r^2 + 2C_1 + C_2 \ , \\ 
G^{2,0}_{N,N}(r,\Delta) = G^{2,0,0}_{N,N}(r,\Delta) &= B_1 r^2 + C_2 \ , \\ 
G^{0,2}_{L,L}(r,\Delta)=G^{0,2,0}_{L,L}(r,\Delta) &= B_2 r^2 + C_2 \ , \\ 
G^{1,1}_{L,N}(r,\Delta)  = G^{1,1,0}_{L,N}(r,\Delta) &= 2B_3 r^2 + C_1 = G^{1,1}_{N,L} \ , \\ 
G^{0,2,0}_{M,M}(r,\Delta) &= C_2  = G^{0,0,2}_{N,N}(r,\Delta) , \\ 
G^{0,1,1}_{N,M}(r,\Delta) &= C_1  \ , \\
G^{0,2,0}_{N,N}(r,\Delta) &= 2C_1+C_2 = 2G^{0,1,1}_{N,M}(r,\Delta) + G^{0,2,0}_{M,M}(r,\Delta) \ ,
\end{align}
where $M$ denotes the second transversal component and the superscripts refer to the number of 
longitudinal, first transverse and second transverse components. After some rearrangement, the 
tensor $G_{ij,lk}$ can be expressed as
\begin{align}
G_{ij,lk} & = (G^{2,0}_{L,L} - G^{2,0}_{N,N} -G^{0,2}_{L,L} -4G^{1,1}_{L,N} + G^{0,2,0}_{N,N}) \frac{r_i r_j r_k r_l}{r^4} \nonumber \\
& \ \ + (G^{1,1}_{L,N}-G^{1,1}_{N,M})(r_i r_k \delta_{jl} + r_i r_l \delta_{jk} + r_k r_j \delta_{il} +  r_j r_l \delta_{ik})/r^2 \nonumber \\
& \ \  + (G^{2,0}_{N,N}-G^{0,2,0}_{M,M}) \frac{r_i r_j}{r^2} \delta_{kl} + (G^{0,2}_{L,L}-G^{0,2,0}_{M,M}) \frac{r_k r_l}{r^2} \delta_{ij} \nonumber \\
&  \ \ + G^{1,1}_{N,M} (\delta_{ik}\delta_{jl} + \delta_{il}\delta_{jk} )  + G^{0,2,0}_{M,M}\delta_{ij}\delta_{kl} \ .
\end{align}
In order to calculate the contribution of this tensor to eq.~\eqref{eq:n3}, 
we must calculate 
\begin{align}
\p_{r_k} G_{\{ijl\}k} = \p_{r_k}(G_{ij,lk} + G_{il,jk} +G_{jl,ik}) \ .
\end{align}
The tensor $G_{\{ijl\}k}$ is now symmetric under the exchange of $i,j,l$, and 
one obtains
\be
\p_{r_k} G_{\{ijl\}k} = 3\left (\partial_r + \frac{2}{r}\right) G^{2,0}_{L,L}(r) 
-\frac{6}{r}(G^{2,0}_{N,N}(r)+G^{1,1}_{L,N}(r)) \ ,
\ee 
where we note that the contributions from the second transversal component cancel out.
The contributions of the correlations tensors between $\tau^\Delta_{ij}$ and velocity-field
increments figuring in the higher-order equations are calculated similarly, resulting 
in 
\be
\p_{r_k} G_{\{i_1\hdots i_n\}k}(r,\Delta) = 
n\left (\partial_r + \frac{2}{r}\right) G^{n-1,0}_{L,L}(r,\Delta)    
-\frac{2n}{r}(G^{n-1,0}_{N,N}(r,\Delta) + G^{n-2,1}_{L,N}(r,\Delta)) \ .
\ee

\subsection{Tensors involving derivatives of velocity field increments}
\label{app:stensors}
In this appendix we derive expressions for the longitudinal components of the 
$S$-and $T$-tensors, which for a general symmetric tensor $\gamma$ have the form 
\begin{align}
S(\gamma)_{\{i_1\hdots i_n\}} & 
= \frac{1}{|S_{n-1}||S_{n-2}|} \sum_{t \in S_{n-1}} \sum_{\sigma \in S_n}
\langle \delta_r \ov_{i_{t(\sigma(1))}} \hdots \delta_r \ov_{i_{t(\sigma(n-2))}}
\partial_{x_k'}\ov_{i_{t(\sigma(n-1))}}' \gamma_{i_{\sigma(n)}k}'\rangle \nonumber \\
&\quad + \frac{1}{|S_{n-1}||S_{n-2}|} \sum_{t \in S_{n-1}} \sum_{\sigma \in S_n}
\langle \delta_r \ov_{i_{t(\sigma(1))}} \hdots \delta_r \ov_{i_{t(\sigma(n-2))}}
\partial_{x_k}\ov_{i_{t(\sigma(n-1))}} \gamma_{i_{\sigma(n)}k}\rangle \ , \\
S(\gamma)_{\{i_1\hdots i_n\}} &
=\frac{1}{|S_{n-1}||S_{n-2}|} \sum_{t \in S_{n-1}} \sum_{\sigma \in S_n}
\langle \delta_r \ov_{i_{t(\sigma(1))}} \hdots \delta_r \ov_{i_{t(\sigma(n-2))}}
\partial_{x_k'}\ov_{i_{t(\sigma(n-1))}}' \gamma_{i_{\sigma(n)}k}\rangle \nonumber \\
&\quad + \frac{1}{|S_{n-1}||S_{n-2}|} \sum_{t \in S_{n-1}} \sum_{\sigma \in S_n}
\langle \delta_r \ov_{i_{t(\sigma(1))}} \hdots \delta_r \ov_{i_{t(\sigma(n-2))}}
\partial_{x_k}\ov_{i_{t(\sigma(n-1))}} \gamma_{i_{\sigma(n)}k}'\rangle 
\ .
\end{align}
Symmetry arguments, i.e. homogeneity and isotropy, restrict the functional form of 
the possible tensors, and one obtains
\begin{align}
\sum_{t \in S_{n-1}} \sum_{\sigma \in S_n} &
\langle \delta_r \ov_{i_{t(\sigma(1))}} \hdots \delta_r \ov_{i_{t(\sigma(n-2))}}
\partial_{x_k}\ov_{i_{t(\sigma(n-1))}} \gamma_{i_{\sigma(n)}k}'\rangle \nonumber \\
&= 
\sum_{t \in S_{n-1}} \sum_{\sigma \in S_n}
\langle \delta_r \ov_{i_{t(\sigma(1))}} \hdots \delta_r \ov_{i_{t(\sigma(n-2))}}
\partial_{x_k'}\ov_{i_{t(\sigma(n-1))}}' \gamma_{i_{\sigma(n)}k}\rangle \ ,  
\end{align}
and 
\begin{align}
\sum_{t \in S_{n-1}} \sum_{\sigma \in S_n} &
\langle \delta_r \ov_{i_{t(\sigma(1))}} \hdots \delta_r \ov_{i_{t(\sigma(n-2))}}
\partial_{x'_k}\ov_{i_{t(\sigma(n-1))}}' \gamma_{i_{\sigma(n)}k}'\rangle \nonumber \\
&= 
\sum_{t \in S_{n-1}} \sum_{\sigma \in S_n}
\langle \delta_r \ov_{i_{t(\sigma(1))}} \hdots \delta_r \ov_{i_{t(\sigma(n-2))}}
\partial_{x_k}\ov_{i_{t(\sigma(n-1))}} \gamma_{i_{\sigma(n)}k}\rangle \ , 
\end{align}
resulting in 
\begin{align}
T_{\{i_i, \hdots , i_n\}} 
= \frac{2}{|S_n||S_{n-1}|} \sum_{t \in S_{n-1}} \sum_{\sigma \in S_n}
\langle \delta_r \ov_{i_{t(\sigma(1))}} \hdots \delta_r \ov_{i_{t(\sigma(n-2))}}
\partial_{x'_k}\ov_{i_{t(\sigma(n-1))}}' \tau^\Delta_{i_{\sigma(n)}k}\rangle \ , 
\end{align}
and 
\begin{align}
S_{\{i_i, \hdots , i_n\}} 
= \frac{2}{|S_n||S_{n-1}|} \sum_{t \in S_{n-1}} \sum_{\sigma \in S_n}
\langle \delta_r \ov_{i_{t(\sigma(1))}} \hdots \delta_r \ov_{i_{t(\sigma(n-2))}}
\partial_{x_k}\ov_{i_{t(\sigma(n-1))}} \tau^\Delta_{i_{\sigma(n)}k}\rangle \ . 
\end{align}

The longitudinal correlation functions 
\begin{align}
S^{n-2}(\gamma) & = \langle (\delta_r \ov_{L})^{n-2} \gamma_{kL}\partial_{k}\ov_L  \rangle \\
T^{n-2}(\gamma) & = \langle (\delta_r \ov_{L})^{n-2} \gamma_{kL}' \partial_{k}\ov_L \rangle \ .
\end{align}
therefore occur with the combinatorial factor $\frac{2n!}{(n-2)!}$ in Eq.~\eqref{eq:hierarchy} for 
$\gamma = \tau^\Delta$.
For $n=2$ we obtain 
\begin{align}
S^0_{L,L}(\gamma) & = \langle \gamma_{kL} \partial_{k}\ov_L \rangle \ , \\
T^0_{L,L}(\gamma) & = \langle \gamma_{kL}' \partial_{k}\ov_L \rangle \ ,
\end{align}
which for $\gamma = \tau$ can also be written as
\begin{align}
\label{eq:stensor_longitudinal}
S^0_{L,L}(\Delta) & = \frac{1}{4}(S_{\{ij\}}(\tau))^0 
= \frac{\partial_r}{4r^4}(r^4 G^{1,0}_{L,L})|_{r=0} 
= \frac{5}{2} \langle s_{LL}\tau^\Delta_{LL}\rangle \ ,\\
\label{eq:ttensor_longitudinal}
T^0_{L,L}(\Delta) & = \frac{1}{4}(T_{\{ij\}}(\tau))^0 
= \frac{\partial_r}{4r^4}(r^4 G^{1,0}_{L,L}) \ . 
\end{align}
The expressions in eq.~\eqref{eq:stensor_longitudinal} and \eqref{eq:ttensor_longitudinal} follow 
from the more general expression
\begin{align}
S_{\{ij\}} &= \partial_{r_k}G_{\{ijk\}}(r,\Delta)|_{\br = \bf 0} \ , \\
T_{\{ij\}} &= \partial_{r_k}G_{\{ijk\}}(r,\Delta)  \ .
\end{align}
\underline{Proof:}\\
The second equation has already been verified in Appendix \ref{app:derivation}, hence it suffices
to consider only the first equation. As shown in Appendix \ref{app:derivation}, the incompressibility 
constraint reduced the divergence of $G_{\{ijk\}}$ to
\begin{align}
\partial_{r_k}G_{\{ijk\}} 
&= 2\partial_{r_k}\left( \langle \ov'_i \tau^\Delta_{jk}\rangle + \langle \ov'_j \tau^\Delta_{ki}\rangle + \langle \ov'_k \tau^\Delta_{ji}\rangle\right) \nonumber \\
&= 2\left( \langle \partial_{r_k} \ov'_i \tau^\Delta_{jk}\rangle + \langle \partial_{r_k}\ov'_j \tau^\Delta_{ki}\rangle + 0 \right) \ ,
\end{align}
where the third term on the RHS vanishes by incompressibility. Hence
\be
\partial_{r_k}G_{\{ijk\}}(r,\Delta)|_{\br = \bf 0} = 
2\left( \langle \partial_{x_k} \ov_i \tau^\Delta_{jk}\rangle + \langle \partial_{x_k}\ov_j \tau^\Delta_{ki}\rangle \right) \ ,
\ee
which coincides with the expression for $S_{\{ij\}}(\Delta)$ given in Eq.~\eqref{eq:stensor_2ndorder} 
\be
S_{\{ij\}}(\Delta) = 2 \langle \partial_{x_k} \ov_i \tau^\Delta_{jk} + \partial_{x_k}\ov_j \tau^\Delta_{ki} \rangle \ .
\ee

\subsection{Tensors involving the pressure}
\label{app:ptensors}
The tensors describing the velocity-pressure correlations are
\begin{align}
P_{\{i_1\hdots i_n \}} 
& = \frac{1}{|S_{n-1}|}\sum_{\sigma \in S_n} 
\langle \delta_r \ov_{i_{\sigma(1)}} \hdots \delta_r \ov_{i_{\sigma(n-1)}} 
\delta_r (\p_k\overline{P}\delta_{k i_{\sigma(n)}}) \rangle 
\nonumber \\
& = \frac{2}{|S_{n-1}|}\sum_{\sigma \in S_n} 
\langle \delta_r\ov_{i_{\sigma(1)}} \hdots \delta_r\ov_{i_{\sigma(n-1)}} 
\p_{r_k}(\overline{P}'+\overline{P}) \delta_{k i_{\sigma(n)}} \rangle 
\nonumber \\
& = \frac{2}{|S_{n-1}|}\sum_{\sigma \in S_n} 
\p_{r_k}\langle \delta_r\ov_{i_{\sigma(1)}} \hdots \delta_r\ov_{i_{\sigma(n-1)}} 
(\overline{P}'+\overline{P}) \delta_{k i_{\sigma(n)}} \rangle 
\nonumber \\
& \ \ - \frac{2}{|S_{n-1}|}\sum_{\sigma \in S_n} 
\left \langle \partial_{r_k}(\delta_r\ov_{i_{\sigma(1)}} \hdots \delta_r\ov_{i_{\sigma(n-1)}}) 
(\overline{P}'+\overline{P})\delta_{k i_{\sigma(n)}} \right \rangle
\ .
\end{align}
The longitudinal components $P^n$ can now be obtained as for the 
tensors $G$, $S$ and $T$:
\begin{align}
P^2 &= 0 \ , \\
P^3 &=  6 \left( \p_r + \frac{2}{r}\right) 
\langle (\delta_r\ov_L)^2(\overline{P}'+\overline{P})\delta_{LL} \rangle
\nonumber \\
& -\frac{12}{r} (\langle (\delta_r\ov_L)^2(\overline{P}'+\overline{P})\delta_{NN} \rangle
 + \langle \delta_r\ov_L \delta_r\ov_N(\overline{P}'+\overline{P})\delta_{LN} \rangle)
\nonumber \\
& \ \ - 3 \left \langle \delta_r\ov_L(\overline{P}'+\overline{P}) 
(\partial_{x_k}\ov_L' + \partial_{x_k}\ov_L)\delta_{kL} \right \rangle
\nonumber \\
&= 6 \p_r \langle (\delta_r\ov_L)^2(\overline{P}'+\overline{P})\rangle
-3 \langle \delta_r\ov_L(\overline{P}'+\overline{P})(s_{LL}' + s_{LL})\rangle \ .
\end{align}

\bibliographystyle{unsrt}
\bibliography{refs,refs2}
\end{document}